\documentclass[english,useAMS,usenatbib]{mn2e}
\usepackage[T1]{fontenc}
\usepackage[utf8]{inputenc}
\usepackage{siunitx}
\usepackage{rotating}
\usepackage{url}
\usepackage{amsmath}
\usepackage{physics}
\usepackage{mathtools}
\usepackage[thinc]{esdiff}
\usepackage{commath}
\usepackage{amssymb}
\usepackage{graphicx}
\usepackage{subfig}
\usepackage{multirow}
\usepackage[table]{xcolor}
\usepackage{pdfpages}
\usepackage{esint}
\usepackage[authoryear]{natbib}
\usepackage{listings}
\usepackage{textcomp}
\usepackage[export]{adjustbox}
\usepackage{booktabs}
\usepackage{ulem}
\usepackage{comment}

\makeatletter

\def\gtsima{$\, \buildrel > \over \sim \,$}
\def\ltsima{$\, \buildrel < \over \sim \,$}
\def\prosima{$\, \buildrel \propto \over \sim \,$}
\def\gsim{\lower.5ex\hbox{\gtsima}}
\def\lsim{\lower.5ex\hbox{\ltsima}}
\def\simgt{\lower.5ex\hbox{\gtsima}}
\def\simlt{\lower.5ex\hbox{\ltsima}}
\def\simpr{\lower.5ex\hbox{\prosima}}
\def\la{\lsim}
\def\ga{\gsim}

\newcommand{\be}{\begin{eqnarray}}
\newcommand{\ee}{\end{eqnarray}}
\def\lsim{\,\lower2truept\hbox{${< \atop\hbox{\raise4truept\hbox{$\sim$}}}$}\,}
\def\gsim{\,\lower2truept\hbox{${> \atop\hbox{\raise4truept\hbox{$\sim$}}}$}\,}

\providecommand{\tabularnewline}{\\}

\title[The first fireworks]{The first fireworks: A roadmap to Population~III stars during the Epoch of Reionization through Pair-Instability Supernovae}
\author[Venditti et al.]{Alessandra Venditti$^{1,2,3,4,5}$\thanks{E-mail:alessandra.venditti@inaf.it}; Volker Bromm$^{5,6}$; Steven L. Finkelstein$^{5}$; Luca Graziani$^{1,2,3}$; \newauthor Raffaella Schneider$^{1,2,3}$ \\
$^{1}$Dipartimento di Fisica, Sapienza, Universit$\grave{a}$ di Roma, Piazzale Aldo Moro 5, 00185, Roma, Italy\\
$^{2}$INFN, Sezione di Roma I, Piazzale Aldo Moro 2, 00185, Roma, Italy\\
$^{3}$INAF-Osservatorio Astronomico di Roma, Via di Frascati 33, 00078, Monte Porzio Catone, Italy\\
$^{4}$Dipartimento di Fisica, Tor Vergata, Universit$\grave{a}$ di Roma, Via Cracovia 50, 00133, Roma, Italy\\
$^{5}$Department of Astronomy, University of Texas at Austin, 2515 Speedway, Stop C1400, Austin, TX 78712, USA \\
$^{6}$Weinberg Institute for Theoretical Physics, University of Texas at Austin, Austin, TX 78712, USA
}

\begin{document}

\date{\today}

\pagerange{\pageref{firstpage}--\pageref{lastpage}} \pubyear{2022}

\maketitle

\label{firstpage}

\begin{abstract}
With the launch of JWST and other scheduled missions aimed at probing the distant Universe, we are entering a new promising era for high-$z$ astronomy. One of our main goals is the detection of the first population of stars (Population III or Pop\ III stars), and models suggest that Pop\ III star formation is allowed well into the Epoch of Reionization (EoR), rendering this an attainable achievement. In this paper, we focus on our chance of detecting massive Pop\ IIIs at the moment of their death as Pair-Instability Supernovae (PISNe).
We estimate the probability of discovering PISNe during the EoR in galaxies with different stellar masses ($7.5 \leq \mathrm{Log}(M_\star/\si{M_\odot}) \leq 10.5$) from six \texttt{dustyGadget} simulations of $50h^{-1}$~cMpc per side. We further assess the expected number of PISNe in surveys with JWST/NIRCam and Roman/WFI. 
On average, less than one PISN is expected in all examined JWST fields at $z \simeq 8$ with $\Delta z = 1$, and O(1) PISN may be found in a $\sim 1$~deg$^2$ Roman field in the best-case scenario, although different assumptions on the Pop\ III IMF and/or Pop\ III star-formation efficiency can decrease this number substantially. Including the contribution from unresolved low-mass halos holds the potential for increased discoveries. 
JWST/NIRCam and Roman/WFI allow the detection of massive-progenitor ($\sim 250 ~ \si{M_\odot}$) PISNe throughout all the optimal F200W-F356W, F277W-F444W, and F158-F213 colors. PISNe are also predominantly located at the outskirts of their hosting haloes, facilitating the disentangling of underlying stellar emission thanks to the spatial-resolution capabilities of the instruments.
\end{abstract}


\section{Introduction}
\label{sec:introduction}

Pair-instability supernovae (PISNe) are a type of supernova (SN) that occurs when a massive star, with a mass between 140\,\si{M_\odot} and 260\,\si{M_\odot}, undergoes a thermonuclear explosion due to the production and subsequent annihilation of electron-positron pairs in its hot core. These reactions lead to a rapid loss of radiation pressure support causing the core to collapse, and thus igniting explosive oxygen and silicon burning that ultimately results in the complete disruption of the star \citep{Rakavy_Shaviv_1967, Barkat_1967, Fraley_1968, Bond_1984, Fryer_2001}. PISNe are particularly interesting because they provide valuable insights on the first generation of stars, known as Population III (Pop III) stars.

Pop III stars formed from pristine gas, almost entirely composed of hydrogen and helium. Because of the absence of heavy elements, they are expected to have very different properties compared to second-generation (Pop II) stars, that formed from the gas pre-enriched by Pop\ III. Most notably, they are expected to be predominantly massive \citep{Abel_2002, Bromm_2002}, with masses ranging from $\sim$10s~\si{M_\odot} to $\sim$100s~\si{M_\odot} \citep{Hosokawa_2011, Hirano_2014, Stacy_2016}, or even $\sim$1000s~\si{M_\odot} \citep{Hirano_2015_UVrad, Hirano_2015_PPS, Susa_2014, Hosokawa_2016, Sugimura_2020, Latif_2022}, implying that some of the most massive Pop III stars may be able to produce PISNe. When a PISN explodes, large amounts of metals are released into the surrounding medium \citep{Heger_Woosley_2002}. Hence, PISNe arising from Pop III stars are thought to be one of the main sources of heavy elements in the early Universe, and are therefore key to understand the process of early chemical enrichment \citep[e.g.][]{Salvadori_2007, Salvadori_2008, Karlsson_2013, deBennassuti_2017, Aguado_2023_PISN-explorer}.

Although a number of potential Pop\ III systems at $z > 6$ have been recently proposed in the literature \citep{Vanzella_2020, Vanzella_2023, Wang_2022, Maiolino_2023}, an unambiguous detection of a Pop\ III-hosting galaxy is currently still missing. The identification of Pop\ III stars in high-$z$ galaxies through different spectroscopic line diagnostics has been extensively discussed over the last decades \citep{Bromm_2001, Inoue_2011, Zackrisson_2011, Mas-Ribas_2016, Nakajima_Maiolino_2022, Trussler_2023, Katz_2023, Cleri_2023}. However, the properties of PISNe, such as their explosive energy output and light curve, can also provide important constraints on the characteristics of their progenitor stars \citep{Mackey_2003, Scannapieco_2005, Whalen_2013, Pan_Loeb_2013, deSouza_2014, Wang_2017}, and thus shed light on the properties of Pop\ III stars. Several simulations have been carried out to model the distinctive light curve and afterglow emission of PISNe arising from Pop IIIs \citep{Scannapieco_2005, Woosley_2007, Pan_2012_PISN, Chen_2014, Whalen_2014, Jerkstrand_2016, Kozyreva_2017, Gilmer_2017, Hartwig_2018}. Simulations suggest that they are very energetic, with total energy produced up to $10^{53}$\,erg, and have an extended light curve duration, $\sim$\,1\,yr in the source frame \citep{Kasen_2011}.

The detection of PISNe is especially challenging due to their rarity. Pop III stars are expected to have formed at high redshifts ($z \sim 20 - 30$; \citealt{Bromm_2013, Klessen_Glover_2023}), when the Universe was only a few hundred million years old, and their rate of formation is expected to have declined rapidly as the Universe became more metal-enriched - although recent simulations suggest a late Pop III star formation, down to the Epoch of Reionization (EoR, $z \sim 6$), is also possible \citep{Xu_2016_latePopIII, Jaacks_2019, Liu_Bromm_2020_2, Sarmento_2018, Sarmento_Scannapieco_2022, Skinner_Wise_2020, Visbal_2020, Venditti_2023}. The exact rate of PISNe across cosmic time is however highly uncertain \citep{Miralda-Escude_Rees_1997, Wise_Abel_2005, Hummel_2012, Pan_2012_PISN, Johnson_2013, Tanaka_2013, Magg_2016}, as it depends on the details of the metal enrichment process, as well as other factors such as the Pop\ III initial mass function (IMF). A more top-heavy Pop\ III IMF, for example, would yield a higher rate of PISNe, as a large fraction of the stars would be massive enough to undergo pair instability \citep{Lazar_Bromm_2022}. However, the exact shape of the Pop III IMF is unconstrained. The typical metallicity of the gas in which Pop III stars formed is also uncertain, as it depends on the details of the primordial star-formation process and the efficiency of metal mixing. Simulations suggest that Pop III stars formed in very low-metallicity environments, with metallicity not higher than $10^{-4} - 10^{-6} ~ \si{Z_\odot}$ \citep{Bromm_2001, Omukai_2005, Maio_2010, Schneider_2012_dustToGas, Schneider_2012_dustSDSSstar, Chiaki_2016, Chiaki_Yoshida_2022}, {although a massive stellar component in excess of what is predicted by a present-day IMF may persist up to metallicity $\lesssim 10^{-2} ~ \si{Z_\odot}$ \citep{Chon_2021, Chon_2022}.

Despite the challenges, several efforts have been made to detect PISNe in the past. One of the most promising candidates proposed is SN 2007bi, a super-luminous SN (SLSN), discovered in 2007 in a nearby dwarf galaxy \citep{Gal-Yam_2008}. SN 2007bi has several properties that are consistent with a PISN, including its high energy output, long duration, measured yield of radioactive nickel and elemental composition of the ejecta. However, its exact nature is still a matter of debate, and other explanations have been proposed in the literature (see e.g. \citealt{Moriya_2010, Dessart_2012}). 
Other examples of SLSNe with slowly evolving light curves at high redshift ($z = 2.05$ and $z = 3.90$) have been found with Keck I follow-up, late-time spectroscopy of two objects first identified in archival data from the Canada-France-Hawaii Telescope Legacy Survey Deep Fields \citep{Cooke_2012}. Possible survey strategies dedicated to the search of PISNe during the EoR with JWST and with the Nancy Grace Roman Space Telescope have been discussed in \citet{Hartwig_2018}, \citet{Regos_2020} and \citet{Moriya_2022_RST}, while the possibility of detecting PISNe in the Euclid Deep Survey (EDS, \citealt{Laureijs_2011, EuclidCollaboration_2022}) at $z \lesssim 2.5$ is studied in \citet{Moriya_2022_Euclid}.

Stellar archaeology can provide a complementary probe to constrain the properties of the first stars - particularly their IMF - and their SN outcomes, by looking at the atmospheres of old and metal-poor stars in our Galaxy. Notably, the detection of a peculiar chemical abundance pattern in a star with [Fe/H$]=-2.5$, possibly indicative of a PISN nucleosynthetic yield \citep{Aoki_2014}, presents evidence that very high masses are indeed allowed for Pop\ III stars \citep{deBennassuti_2017}. Recently, \citet{Xing_2023} have reported the identification of a halo star, with [Fe/H$]=-2.42$, whose abundance pattern is remarkably well fit by a PISN imprint from a 260\,\si{M_\odot} progenitor, although a core-collapse SN-origin is at least equally plausible with the currently measured elements \citep{Jeena_2023}. Other instances of metal-poor stars with suggested PISN signatures in their atmosphere have been discussed in \citet{Salvadori_2019} and \citet{Aguado_2023_PISN-explorer}, and a contribution of PISNe from massive Pop\ III stars was also suggested to explain the unusual abundance feature of Fe and Mg in the broad-line region of a quasar at $z = 7.54$ \citep{Yoshii_2022}. Simulations of early chemical enrichment \citep[e.g.][]{Jeon_2015} have suggested that PISN-produced material predominantly resides in the intergalactic medium (IGM), raising the possibility of using the ultra-luminous afterglows of high-redshift gamma ray bursts (GRBs) to probe this IGM signature with deep absorption spectroscopy \citep{Wang_2012}.

An important factor in the detectability of PISNe is their host environment. Pop\ III stars are expected to have formed predominantly in low-mass, faint haloes. However, some models also suggest that Pop\ III could still form in more massive and luminous haloes \citep{Liu_Bromm_2020_2, Bennet_Sijacki_2020, Riaz_2022, Venditti_2023}. Although massive galaxies are generally easier to detect, their higher luminosity might even hinder our ability to clearly identify a PISN in such hosts. This is especially true if the SN is observed at late times, when the emission has faded considerably \citep{Kasen_2011}. An extensive study of the expected rate of PISNe as a function of their host environments is currently missing.

This work is aimed at assessing the probability of observing PISNe arising from Pop\ III stars during the EoR, in galaxies with different stellar masses. We rely on a suite of six $50h^{-1} ~ \si{cMpc}$ simulations already introduced in \citet{DiCesare_2022} and \citet{Venditti_2023}, carried out with the hydrodynamical code \texttt{dustyGadget} \citep{Graziani_2020}. Our findings can guide the design of optimal strategies to search for potential PISN-hosting candidates, both from archival data and from dedicated surveys. While achieving the first clear direct detection of a PISN event would be a remarkable accomplishment in and of itself, finding PISNe at high redshifts would also have deeper implications for the search of the first stars. As PISNe are in fact the natural outcome of very massive stars, hardly attainable for normal Pop\ II/I stars, they can be used as markers to identify potential Pop\ III hosts. Hence, the synergy between wide-field PISNe searches and spatially resolved observations of their host galaxies (possibly enhanced by gravitational lensing) may offer one of our finest tools to directly probe Pop\ III stars and test our predictions on primordial star-formation.

The paper is organised as follows. Section~\ref{sec:method} presents our methodology: Section~\ref{sec:method_simulations} describes our simulation suite, while Section~\ref{sec:method_probPISN} outlines our strategy for evaluating the probability of finding PISNe during the EoR from the simulations. Our results are presented in Section~\ref{sec:results}. Specifically, in Section~\ref{sec:results_PISNNumber} we compute the expected number of PISNe in galaxies of different stellar mass per unit halo and volume, and we predict the expected number of PISNe within JWST and Roman fields. In Section~\ref{sec:results_PISNeVSStars} we discuss our capability of observing PISNe and discriminating from the stellar emission of their host galaxies, by considering their integrated emission (Section~\ref{sec:results_PISNeVSStars_integratedEmission}) and possibly resolved observations (Section~\ref{sec:results_PISNeVSStars_resolvedObservations}). A critical discussion is presented in Section~\ref{sec:discussion}, including a number of caveats in our modelling (Section~\ref{sec:discussion_caveats}), and a discussion on the implication of either future detections or non-detections of PISNe (Section~\ref{sec:discussion_futureDetections}), and on the advantages and disadvantages of many possible detection strategies (Section~\ref{sec:discussion_detectionStrategies}). Finally, we summarize and offer conclusions in Section~\ref{sec:conclusions}.

\section{Methodology}
\label{sec:method}

\subsection{Simulating the cosmological environment}
\label{sec:method_simulations}

Our study is based on a suite of eight cosmological simulations carried out with the hydrodynamical code \texttt{dustyGadget}. The code is extensively described in \citet{Graziani_2020}, especially regarding its most novel feature, i.e. the implementation of a self-consistent model for dust production and evolution, while our simulations (hereafter named as U6 - U13) are introduced in \citet{DiCesare_2022} and \citet{Venditti_2023}. More specifically, \citet{DiCesare_2022} studied the main scaling relations in the context of available observations and model predictions, while \citet{Venditti_2023} focused on the properties of galaxies hosting Pop\ III star formation during the EoR, which are of particular interest for the present work. We briefly describe relevant aspects of the simulations here, and refer to the aforementioned papers for further details.

All the simulated volumes have a (comoving) size of $50h^{-1} ~ \si{cMpc}$, a total number of $2 \times 672^3$ particles and a mass resolution for dark matter/gas particles of $3.53 \times 10^7 h^{-1} ~ \si{M_\odot}$/$5.56 \times 10^6 h^{-1} ~ \si{M_\odot}$. They assume a $\mathrm{\Lambda}$CDM cosmology, consistent with \citet{Planck_2015} parameters ($\Omega_{\mathrm{m,0}} = 0.3089$, $\Omega_{\mathrm{b,0}} = 0.0486$, $\Omega_{\mathrm{\Lambda},0} = 0.6911$, $h = 0.6774$), and share a common feedback setup based on \citet{Graziani_2020}. A cold gas phase density threshold for star formation of $n_\mathrm{th} \simeq 300 ~ \si{cm^{-3}}$ is adopted. When this threshold is reached, stellar populations represented by stellar particles of mass $\sim 2 \times 10^6 ~ \si{M_\odot}$ are generated in a single, instantaneous burst. Their IMF/metal yields (following C, O, Mg, S, Si and Fe) are assigned according to their metallicity $Z_\star$, given a critical metallicity $Z_\mathrm{crit} = 10^{-4} ~ \si{Z_\odot}$\footnote{We assume $Z_\mathrm{\odot} = 0.02$ \citep{Anders_Grevesse_1989}.} \citep{Tornatore_2007_PopIII, Maio_2010, Graziani_2020}, specifically:
\begin{enumerate}
    \item a Salpeter-like IMF in the mass range [100, 500] \si{M_\odot} for Pop\ III stars ($Z_\star < Z_\mathrm{crit}$), with mass-dependent yields describing the metal pollution from stars in the PISN range [140, 260] \si{M_\odot} \citep{Heger_Woosley_2002};
    \item a standard Salpeter IMF \citep{Salpeter_1955} in the mass range [0.1, 100] \si{M_\odot} for Pop\ II stars ($Z_\star \geq Z_\mathrm{crit}$), with mass and metallicity-dependent yields describing the metal pollution from long-lived, low/intermediate-mass stars \citep{vanDenHoek_Groenewegen_1997}, high mass stars (> 8 \si{M_\odot}), dying as core-collapse SNe \citep{Woosley_Weaver_1995}, and SNeIa \citep{Thielemann_2003}.
\end{enumerate}
Pop\ II/I stars with masses $\geq 40 \; \si{M_\odot}$ and Pop\ III stars outside the PISN mass range do not contribute to the metal enrichment, as they are assumed to directly collapse into black holes (not followed explicitly here). The impact of our choice of the Pop\ III IMF, and hence of neglecting the contribution of traditional core-collapse SNe from Pop\ IIIs, is discussed in Section \ref{sec:discussion}.

The gas chemical evolution model is adopted from \citet{Tornatore_2007_chemicalFeedback}. Dust and metals are spread in the ISM through a spline kernel, and galactic winds are also modelled as in \citet{Springel_Hernquist_2003} with a constant velocity of $500 ~ \si{km.s^{-1}}$, as indicated by observations of normal galaxies in the ALMA Large Program to INvestigate [CII] at Early times (ALPINE) survey \citep{Ginolfi_2020}.

The identification of dark matter haloes and their substructures is performed in post-processing with the \texttt{AMIGA} halo finder \citep{Knollmann_Knebe_2009}. Finally, the simulations are carried out from $z \simeq 100$ down to $ z \simeq 4$, while this work only focuses on the redshift range $6 \lesssim z \lesssim 8$. Indeed, the number of well-resolved Pop\ III-hosting haloes (with a stellar mass $\mathrm{log} (M_\star/ \si{M_\odot}) \gtrsim 7.5$, corresponding to a number of stellar particles $\gtrsim 20$) decreases significantly at higher redshifts, while a drop in Pop\ III star formation is expected at lower redshifts due to the joint effect of cosmic metal enrichment and ultraviolet (UV)/Lyman-Werner (LW) radiative feedback. A more extensive discussion can be found in \citet{Venditti_2023}.

The simulations described in \citet{DiCesare_2022} and \citet{Venditti_2023} have been chosen for the present study because, given the rarity of PISNe, having a big simulated volume is key to capture the statistics of these events. The convergence of our model has been tested in \citet{Graziani_2020}\footnote{While \citet{Graziani_2020} mainly focused on the results for the dust content of the most massive and resolved galaxies, we find that the uncertainty on the Pop\ III star-formation-rate density arising from numerical effects is comparable in magnitude to the uncertainty resulting from different choices of the initial conditions, at the redshifts of interest.}. Given our limited mass resolution, we focus on a much higher mass regime for Pop\ III-hosting halos ($\mathrm{log} (M_\star/ \si{M_\odot}) \gtrsim 7.5$) than that of the first mini-halos. In a future work, zoom-in simulations will be performed to investigate the contribution of low-mass halos to Pop\ III star formation across cosmic times, including the very first star-forming regions.

\subsection{Computing the probability of finding PISNe}
\label{sec:method_probPISN}

The average number of PISNe, $\overline{N}_\mathrm{PISN}$, produced by a Pop III stellar population per unit Pop III mass $M_\mathrm{III}$, with our assumed IMF $\phi(m)$ (whose lower and upper limits are $m_\mathrm{low}$ and $m_\mathrm{up}$, respectively), is given by:
\begin{equation}
    \frac{\overline{N}_\mathrm{PISN}}{M_\mathrm{III}} = \frac{\int_{140 ~ \si{M_\odot}}^{260 ~ \si{M_\odot}} \phi(m) \dd{m}}{\int_{m_\mathrm{low}}^{m_\mathrm{up}} m \phi(m) \dd{m}} \simeq 0.0022 ~ \si{M_\odot^{-1},}
    \label{eq:avNumberOfPISNePerUnitMass}
\end{equation}
indicating an average of one PISN event per 460~\si{M_\odot}. For Pop\ III stellar populations with mass $\sim 2 \times 10^6 ~\si{M_\odot}$, this would result in more than 4000 PISNe produced on average by each stellar population (but see the discussion below on more realistic outcomes).

\begin{figure}
    \centering
    \includegraphics[width=\linewidth]{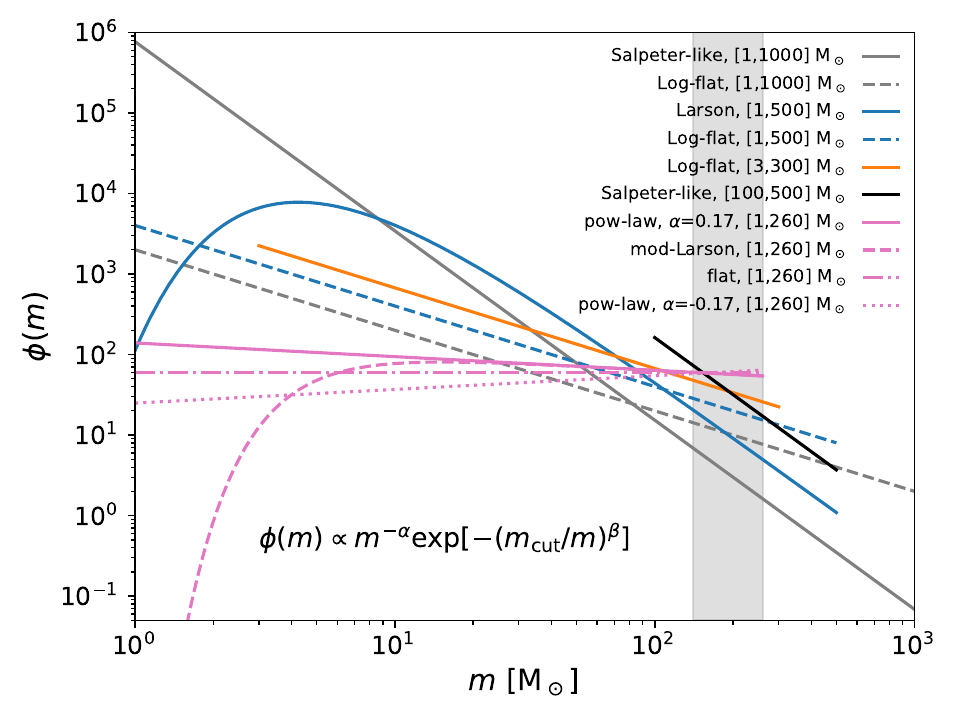}
    \caption{Compilation of various theoretical models for the Pop III IMF, $\phi(m)$, normalised to a total mass of $2 \times 10^6$~\si{M_\odot} (i.e. approximately our Pop III stellar particle mass resolution). The models shown here are from \citet{Schauer_2022} (\textit{blue}), \citet{Hartwig_2018} (\textit{orange}) and \citet{Jaacks_2018_metalEnrichment} (\textit{pink}), and they can all be expressed through the general expression at the bottom (Equation~\ref{eq:popIIIIMFs}), with parameters specified in Table~\ref{tab:popIIIIMFs}. Our Salpeter-like IMF \citep{Graziani_2020, DiCesare_2022, Venditti_2023} is shown as a \textit{black, solid line}. We also include IMFs with a broader mass range, extending up to 1000~\si{M_\odot}, as allowed from ab-initio simulations of star-forming clouds (e.g. \citealt{Susa_2014, Hirano_2015_UVrad, Hosokawa_2016}, \textit{grey}). The \textit{gray shaded region} indicates the stellar progenitor mass range of PISNe.}
    \label{fig:popIIIIMFs}
\end{figure}
\begin{table}
    \centering
    \caption{Values of the average number of PISNe produced per unit Pop III mass $\overline{N}_\mathrm{PISN} / M_\mathrm{III}$ computed for a compilation of Pop III IMFs, expressed through the general form of Equation~\ref{eq:popIIIIMFs}, with parameters $\alpha$, $\beta$, $m_\mathrm{cut}$ and mass range $m_\mathrm{range}$ as specified. The IMF adopted in this work is highlighted in \textit{bold}. See Figure~\ref{fig:popIIIIMFs} for references.}
    \begin{tabular}{c|c|c|c|c|c}
         & $\alpha$ & $\beta$ & $m_\mathrm{cut}$ & $m_\mathrm{range}$ & $\overline{N}_\mathrm{PISN} / M_\mathrm{III}$ \\
         & & & [\si{M_\odot}] & [\si{M_\odot}] & [\si{10^{-3}~M_\odot^{-1}}] \\
         \hline
         Salpeter-like & 2.35 & 0 & 0 & [1, 1000] & 0.20 \\
         Log-flat & 1 & 0 & 0 & [1,1000] & 0.62 \\
         Larson & 2.35 & 1 & 10 & [1500] & 0.62 \\
         Log-flat & 1 & 0 & 0 & [1,500] & 1.24 \\
         Log-flat & 1 & 0 & 0 & [3,300] & 2.08 \\
         \textbf{Salpeter-like} & \textbf{2.35} & \textbf{0} & \textbf{0} & \textbf{[100,500]} & \textbf{2.17} \\
         pow-law & 0.17 & 0 & 0 & [1,260]$^{\dagger}$ & 3.41 \\
         mod-Larson & 0.17 & 2 & 4.47 & [1,260]$^{\dagger}$ & 3.42 \\
         flat & 0 & 0 & 0 & [1,260]$^{\dagger}$ & 3.55 \\
         pow-law & -0.17 & 0 & 0 & [1,260]$^{\dagger}$ & 3.68 \\
    \end{tabular}
    $^{\dagger}$The mass range for these IMFs has been extended up to 260~\si{M_\odot} with respect to the original upper limit (150~\si{M_\odot}) of \citet{Jaacks_2018_metalEnrichment}, to account for the proposed constraint on the Pop III IMF upper limit from \citet{Xing_2023}. We note that higher values for $\overline{N}_\mathrm{PISN} / M_\mathrm{III}$ (in the range $9.32 - 12.13$) are found when this constraint is removed.
    \label{tab:popIIIIMFs}
\end{table}
We note that $\overline{N}_\mathrm{PISN} / M_\mathrm{III}$ can vary depending on the adopted IMF. As both the shape and mass range of the Pop III IMF are still largely uncertain, in Table~\ref{tab:popIIIIMFs} we evaluate this quantity for a compilation of IMFs that have been adopted in the literature to describe Pop III stellar populations (see Figure~\ref{fig:popIIIIMFs}). The shape of all these IMFs can be expressed in the general form \citep{Lazar_Bromm_2022}:
\begin{equation}
    \phi(m) \propto m^{-\alpha} \exp[-(m_\mathrm{cut} / m)^\beta],
    \label{eq:popIIIIMFs}
\end{equation}
i.e. a modified-Larson IMF \citep{Larson_1998}, with various parameters describing the slope ($\alpha$) and the exponential cut-off at low stellar masses ($m_\mathrm{cut}$, $\beta$). We note that $\overline{N}_\mathrm{PISN} / M_\mathrm{III}$ can vary in the range $\sim [0.2 - 4] \times 10^{-3} ~ \si{M_\odot^{-1}}$ over all the included IMFs.

Another source of uncertainty comes from our Pop\ III stellar mass resolution element $M_\mathrm{III,res} \sim 2 \times 10^6 ~ \si{M_\odot}$. Indeed, since the expected number of PISNe is directly proportional to the total Pop\ III mass $M_\mathrm{III}$, it is important to carefully estimate the amount of stellar mass produced in a single star-formation event. In principle, this could be lower than $M_\mathrm{III,res}$, which should more accurately be interpreted as the amount of extremely metal-poor gas above our density threshold that is available for star formation. We can formally express this through an efficiency factor $\eta_\mathrm{III} < 1$:
\begin{equation}
    M_\mathrm{III} = \eta_\mathrm{III} M_\mathrm{III,res},
    \label{eq:popIIIResolutionElement}
\end{equation}
and place a lower limit on $\eta_\mathrm{III} \sim 0.01$ from a wealth of simulation data describing the first sites of Pop III star formation in mini-haloes (see \citealt{Bromm_2013} and references therein). However, mini-haloes at the Rees-Ostriker cooling threshold \citep{Rees_Ostriker_1977} are known to be very inefficient hosts for star formation, given their shallow potential well. We can thus argue that $\eta_\mathrm{III}$ could be larger in more massive haloes at later times, and explore our results for a range of different values of $\eta_\mathrm{III} \geq 0.01$. As no available simulation constrains the mass regime we are currently exploring, we allow this parameter to vary up to about an order of magnitude more than our lower limit $\eta_\mathrm{III} = 0.01$. An empirical hint that higher values of $\eta_\mathrm{III}$ (even up to $\eta_\mathrm{III} \simeq 0.3$) might be allowed at least in some cases is offered by the tentative detection of HeII emission at $\simeq 2.5 ~ \si{kpc}$ from an exceptionally luminous galaxy at $z = 10.6$, that might indicate the presence of a Pop\ III cluster with a top-heavy IMF and a total Pop\ III mass of $6 - 7 \times 10^5 ~ \si{M_\odot}$ \citep{Maiolino_2023}. With this caveat in mind, the average number of PISNe produced by a single Pop\ III stellar population is reduced to $\sim 4000 \, \eta_\mathrm{III}$, for our adopted IMF.

We emphasize that, although both the Pop\ III IMF and star-formation efficiency are still very uncertain, they only affect the expected number of PISNe through the scaling parameters $\overline{N}_\mathrm{PISN}/M_\mathrm{III}$ and $\eta_\mathrm{III}$. Hence, our results can be easily adjusted if further constraints on either quantities will become available. In fact, our goal is to provide a framework to estimate the probability of achieving a direct detection of PISNe and using these PISN discoveries to tag sites of late Pop~III star formation close to reionization, rather than compute an exact rate. In the following, the case $\overline{N}_\mathrm{PISN}/M_\mathrm{III} = 2.17 \times 10^{-3} ~ \si{M_\odot^{-1}}$ (corresponding to our Salpeter-like IMF in the range [100, 500]~\si{M_\odot}) and $\eta_\mathrm{III} = 0.1$ will be referred to as our ``reference model'', as baseline for our discussion. We emphasize that this choice of reference values is for convenience of presentation only, and does not imply that they are the most realistic ones.

\begin{figure}
    \centering
    \includegraphics[width=\linewidth]{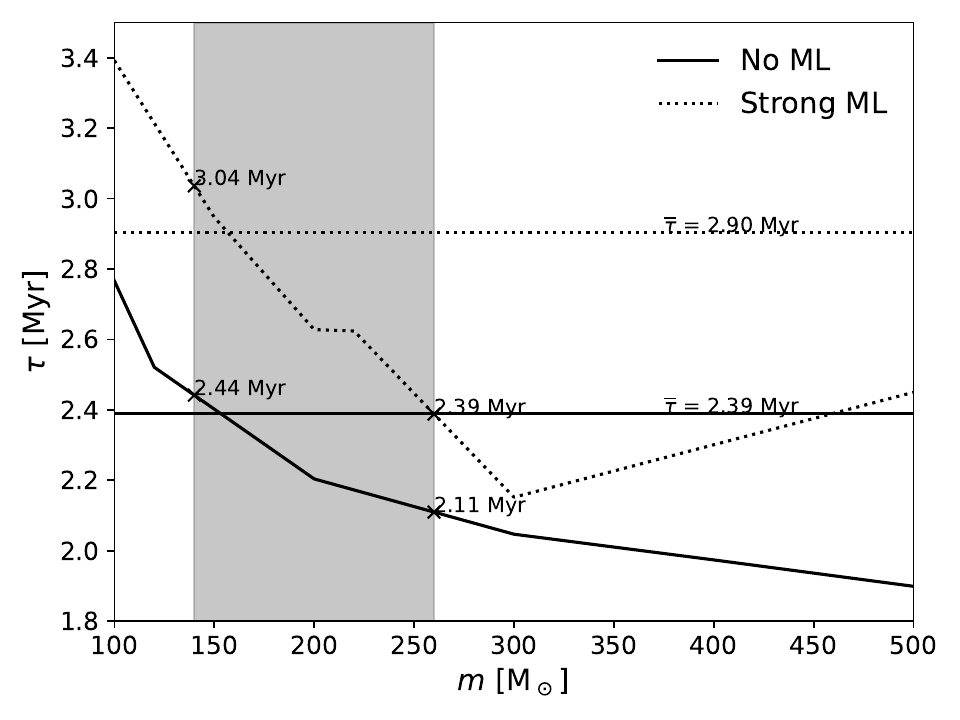}
    \caption{Pop III main-sequence lifetimes $\tau$ as a function of the initial mass $m$ of the stars, from \citet{Schaerer_2002}, in the mass range of our assumed IMF. The \textit{black, solid line} refers to the model assuming no mass loss (see table~4 of the original paper), while the \textit{black, dotted line} refers to the model assuming strong mass loss (see their table~5). The interpolated lifetimes at the limits of the PISN progenitors mass range [140, 260]~\si{M_\odot} (\textit{grey, shaded area}) are indicated for the two models. The average lifetimes $\overline{\tau}$ of Pop III stellar populations for the two models with our assumed IMF are also shown, on top of the \textit{horizontal, solid/dotted, black lines}.}
    \label{fig:popIIILifetimes}
\end{figure}

Since Pop\ III stellar particles in our simulations are formed in an instantaneous burst, and Pop\ IIIs with initial masses $m$ in the range [140, 260]~\si{M_\odot} explode as PISNe over a time span $\Delta t_\mathrm{PISN} = \tau(140 ~ \si{M_\odot}) - \tau(260 ~ \si{M_\odot})$, where $\tau(m)$ is the lifetime of a star with initial mass $m$, we can approximately compute the average number of PISNe expected to be seen at a given time $t$ within $\Delta t_\mathrm{PISN}$ as:
\begin{equation}
    \overline{N}_\mathrm{PISN, t} \simeq \overline{N}_\mathrm{PISN} \times \frac{\Delta t_\mathrm{prompt}}{\Delta t_\mathrm{PISN}},
    \label{eq:avNumberOfPISNe_atFixedTime}
\end{equation}
where $\Delta t_\mathrm{prompt} \simeq$~1~yr is the rest-frame time interval over which the prompt PISN emission is expected to be bright \citep{Kasen_2011}.

The Pop\ III lifetimes $\tau(m)$ as a function of initial stellar mass $m$ are taken from the models of \citet{Schaerer_2002}\footnote{Note that these are in fact main-sequence lifetimes. However, \citet{Marigo_2001} show that the He burning lifetime is $\lesssim 10 \%$ of the main sequence phase.}. Given the uncertainties for existing Pop\ III evolution models, they reported results for models assuming strong mass loss\footnote{As hot-star winds are mainly driven by photon momentum transfer through metal absorption, any mass loss arising from metal-poor stars is in fact expected to be sub-dominant. This is confirmed by observations of wind velocities and average momenta in the Magellanic clouds \citep{Kudritzki_Puls_2000} and by theoretical studies \citep{Baraffe_2001}. Nonetheless, e.g. \citet{Regos_2020} considered models for Pop\ IIIs that have lost their outer envelope due to rotationally-induced mixing before exploding as PISNe.} arising from high-mass Pop\ III stars and models assuming no mass loss at all (see their tables~5 and 4 respectively). In Figure~\ref{fig:popIIILifetimes}, we show $\tau(m)$ for both models over the mass range of our assumed Pop\ III IMF, to emphasize the existing differences, indicating that mass loss can result in longer lifetimes by up to a factor of $\simeq 1.3$, and even in an inversion of the trend for masses $m \gtrsim 300 ~ \si{M_\odot}$, with $\tau$ increasing rather than decreasing for larger stellar masses. The lifetimes of stars with masses 140~\si{M_\odot} and 260~\si{M_\odot} are also shown in the plots. With these lifetimes, $\Delta t_\mathrm{PISN} \simeq 0.65/0.33$~Myr respectively for the models assuming strong/no mass loss. This results in an average number of PISNe from our stellar populations of the order of $\sim 10^{-2} \, \eta_\mathrm{III}$ at a given time\footnote{Note that for the strong-mass-loss case this is likely an upper limit, as a fraction of the stars will not explode as PISN if they experience strong mass loss due to rotation.} (Equation~\ref{eq:avNumberOfPISNe_atFixedTime}).

To compute the average number of PISNe that we expect to find in haloes with stellar mass in the range [$M_\star$, $M_\star + \Delta M_\star$] of a given simulated volume $V$ and simulation snapshot, i.e. at a given time $t$, we sum over all Pop\ III particles with age $\tau(260~\si{M_\odot}) \leq t_\mathrm{III} \leq \tau(140~\si{M_\odot})$ in these haloes to recover the total mass $M_\mathrm{III,PISN} (M_\star)$ of Pop\ III stellar populations that can produce PISNe within a time $\Delta t_\mathrm{PISN}$. The average number of PISNe as a function of $M_\star$ is then:
\begin{equation}
     \overline{N}_\mathrm{PISN, t} (M_\star) \simeq M_\mathrm{III,PISN} (M_\star) \times \frac{\overline{N}_\mathrm{PISN}}{M_\mathrm{III}} \times \frac{\Delta t_\mathrm{prompt}}{\Delta t_\mathrm{PISN}}.
\end{equation}

If we assume the actual number of PISNe $k$ found in a fixed volume $V$ at time $t$ follows a Poisson distribution $P(k; \lambda) = \lambda^k e^{-\lambda} / k!$ with parameter $\lambda \equiv \overline{N}_\mathrm{PISN, t} (M_\star)$, we can estimate the probability of finding at least 1 PISN in haloes with stellar mass within [$M_\star$, $M_\star + \Delta M_\star$] as:
\begin{equation}
    P(\geq \mathrm{1 ~ PISN}) = 1 - P(0; \lambda) = 1 - e^{-\lambda},
\end{equation}
and if the average number of PISNe that we find in the volume is indeed very small - as we would expect - we will have $P(\geq \mathrm{1 ~ PISN}) \simeq \lambda \equiv \overline{N}_\mathrm{PISN, t} (M_\star)$.

\section{Results}
\label{sec:results}

\subsection{Expected number of PISNe in galaxies of different mass}
\label{sec:results_PISNNumber}

\begin{figure*}
    \centering
    \includegraphics[width=\linewidth]{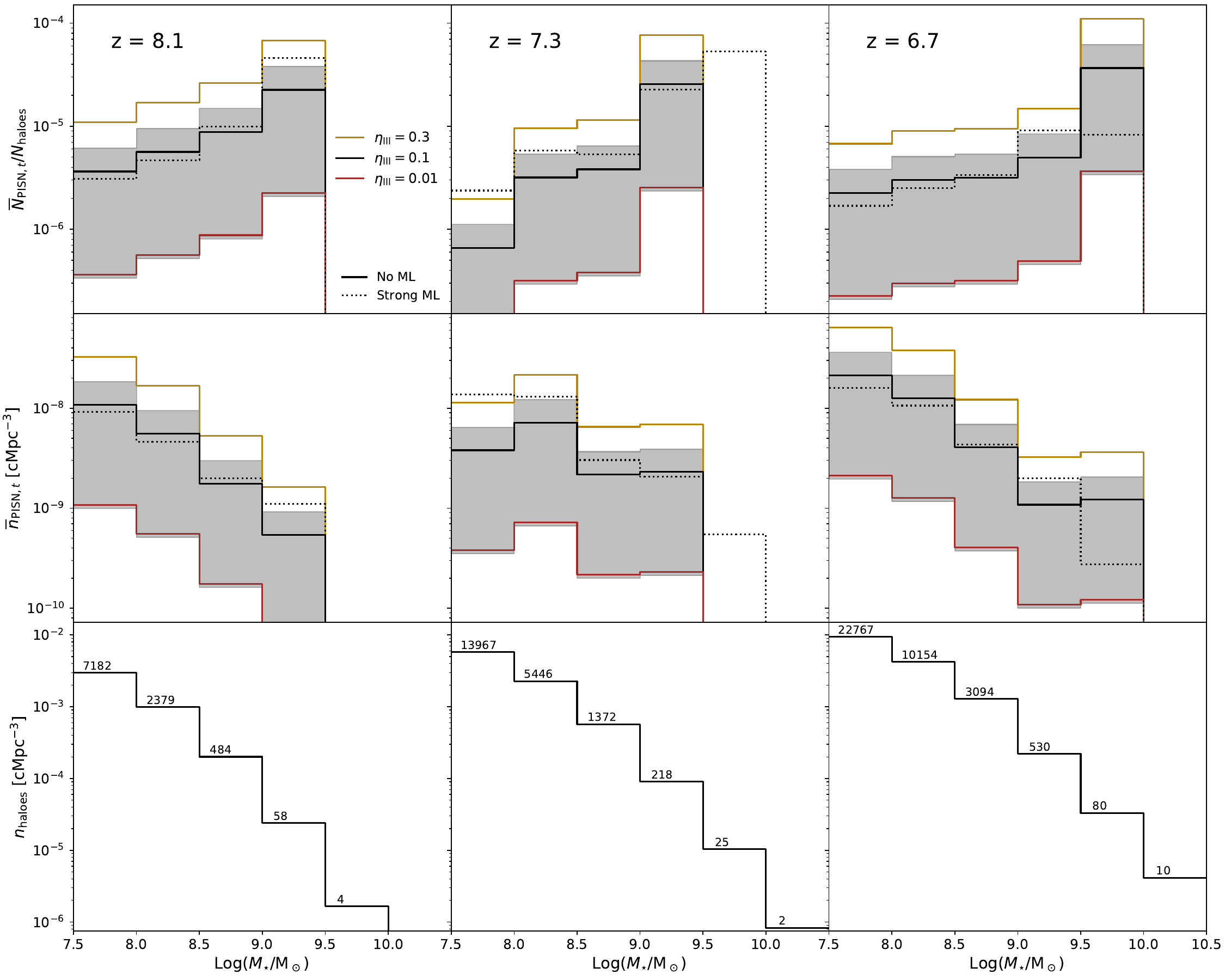}
    \caption{\textbf{Top/middle panels:} Average number of PISNe per halo ($\overline{N}_\mathrm{PISN,t} / N_\mathrm{haloes}$)/per unit volume ($\overline{n}_\mathrm{PISN,t}$) that we expect to find at a given redshift $z$ in haloes within a given range of stellar mass $M_\star$ as a function of $M_\star$, computed in six bins of $M_\star$ (with a spacing of 0.5~dex in the range $7.5 \leq \mathrm{Log} M_\star / \si{M_\odot} < 10.5$). The various colours in the top panels refer to different values of the efficiency $\eta_\mathrm{III}$ of a single star-formation episode; $\eta_\mathrm{III} = 0.01$ (\textit{gold}), $\eta_\mathrm{III} = 0.1$ (\textit{black}), and $\eta_\mathrm{III} = 0.3$ (\textit{brown}). The \textit{solid lines} refer to the model assuming no mass loss, while the \textit{dotted line} refers to the model assuming strong mass loss for the case $\eta_\mathrm{III} = 0.1$ (Figure~\ref{fig:popIIILifetimes}). The \textit{shaded area} also demonstrates the expected scatter around the model assuming no mass loss and $\eta_\mathrm{III} = 0.1$ for different assumed IMFs (see Figure~\ref{fig:popIIILifetimes} and Table~\ref{tab:popIIIIMFs}). \textbf{Bottom panels}: comoving number density $n_\mathrm{haloes}$ of haloes found in each $M_\star$ bin. The total number of haloes found in each bin is also indicated on top of the bin. Results are shown for the combined simulated volumes U6, U7, U8, U10, U12, and U13 at redshifts $z = 8.1$ (\textbf{left panels}), $z = 7.3$ (\textbf{middle panels}) and $z = 6.7$ (\textbf{right panels}).}
    \label{fig:PISN}
\end{figure*}

We compute the average number of PISNe at a given redshift $z$ as a function of the stellar mass $M_\star$ of the halo in which they are found, using the method outlined in Section~\ref{sec:method_probPISN} (see Equation~\ref{eq:avNumberOfPISNe_atFixedTime}). We consider Pop\ III stellar particles within all the simulated volumes U6, U7, U8, U10, U12, and U13 to provide the largest possible statistics\footnote{Data from U9 and U11 is not included as the snapshot dumps for these cubes are not aligned with the others.}, meaning we are virtually probing a total volume $V_\mathrm{eff} = 6 \times (50h^{-1} ~ \si{cMpc})^3$. We also focused on six bins of $M_\star$ spaced by 0.5~dex in the range $7.5 \leq \mathrm{Log} M_\star / \si{M_\odot} < 10.5$, for three different redshift points, $z = 8.1, \, 7.3$ and 6.7. 

The results are shown in the top and middle panels of Figure~\ref{fig:PISN}, for the models assuming no/strong mass loss (solid/dotted, black lines). In the plots, the average number of PISNe is normalised by the total number of haloes $N_\mathrm{haloes}$ in each bin (top panels) and by the total volume $V_\mathrm{eff}$ (middle panels), to recover the expected number of PISNe per halo/per unit volume; these numbers provide an indication respectively on the number of galaxies one should be looking at to have a reasonable chance of observing PISNe at a given redshift and on the average comoving number density of PISNe expected in a blind survey. Different assumptions on the efficiency $\eta_\mathrm{III}$ (see Equation~\ref{eq:popIIIResolutionElement}) and on the IMF $\phi(m)$ (see Equation~\ref{eq:avNumberOfPISNePerUnitMass}) are explored. The total number and comoving number density of haloes are also shown in the bottom panels to demonstrate our explored statistics, and we refer to \citet{DiCesare_2022} for a detailed comparison of the \texttt{dustyGadget} galaxy stellar mass functions at various redshifts with observations and with other models and simulations.

We see that the probability of finding PISNe in our simulated volumes is small but non-negligible: in our reference model for Pop\ III (Salpeter-like IMF in the range [100, 500]~\si{M_\odot}, Pop\ III star formation efficiency of $\eta_\mathrm{III} = 0.1$ and no mass loss), the comoving number density of PISNe can reach values $\sim 10^{-1} ~ \si{cMpc^{-3}}$, while the average number of PISNe per halo can be up to $\sim 5 \times 10^{-6}$, meaning we would expect about 1 PISN every two hundred thousand haloes, on average (if we take into account the contributions of all the considered stellar mass bins).
These numbers depend however on the considered redshift, on the assumed Pop\ III model and on the halo stellar mass. Particularly, high-mass haloes ($10^{9.5} ~ \si{M_\odot} \lesssim M_\star \lesssim 10^{10} ~ \si{M_\odot}$) at $z = 6.7$ seem to be the most favourable for the search of PISNe at high redshifts ($\overline{N}_\mathrm{PISN} / N_\mathrm{haloes} \gtrsim 3 \times 10^{-5}$ in our reference model, see top panels). In general, the probability of finding PISNe in a given halo increases with $M_\star$, dropping to zero at the highest masses ($M_\star \gtrsim 10^{9.5} ~ \si{M_\odot}$ at $z = 8.1$, $M_\star \gtrsim 10^{10} ~ \si{M_\odot}$ at $z = 7.3/6.7$). On the other hand, given the higher number of haloes at the low-mass end, the expected number density of PISNe in lower-mass galaxies is generally higher, even though the fraction of low-mass galaxies hosting PISNe is lower; the highest number density is found in haloes with $M_\star \lesssim 10^8 ~ \si{M_\odot}$ at $z = 6.7$ ($\overline{n}_\mathrm{PISN} \sim 2 \times 10^{-8} ~ \si{cMpc^{-3}}$ in our reference model, see middle panels). We emphasize that the statistics for the highest stellar mass bins is very limited, as no more than a few/a few tens of haloes are found in these bins (see bottom panels), and hence our results are very sensitive to the adopted selection criterion, see e.g. the difference between the no/strong mass loss cases.

While the discrepancies due to the assumption of no/strong mass loss in our estimate of Pop III lifetimes are mostly sub-dominant (apart from the rare massive haloes), a more significant variation is expected when considering different IMFs and/or different values of $\eta_\mathrm{III}$. In particular, a factor of ten in the assumed $\eta_\mathrm{III}$ yields a difference of $\sim 1$~dex in the number of PISNe. The scatter resulting from different assumptions on the IMF (covering all the cases explored in Table~\ref{tab:popIIIIMFs} and shown as a shaded region for the reference case $\eta_\mathrm{III} = 0.1$), is instead slightly higher than 1~dex.

\begin{figure}
    \centering
    \includegraphics[width=\linewidth]{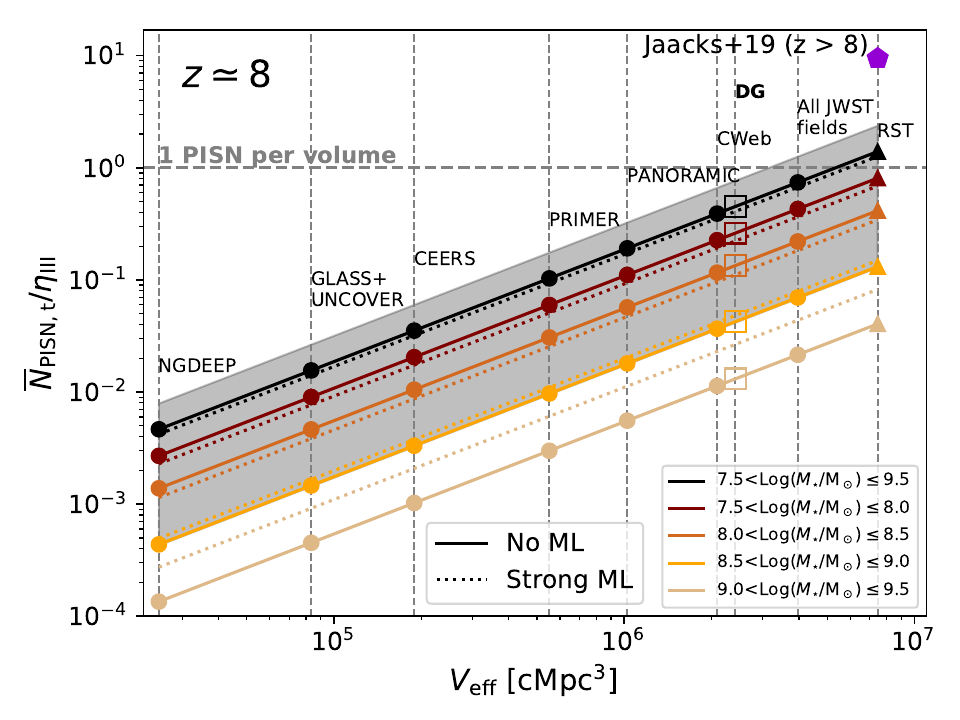}
    \caption{Average number of PISNe $\overline{N}_\mathrm{PISN,t}$ (normalised to the Pop\ III formation efficiency $\eta_\mathrm{III}$) as a function of the effective survey volume $V_\mathrm{eff}$ at $z = 8.1$. \textit{Vertical, dashed lines} indicate the effective volume of selected JWST surveys and their cumulative volume at $z \simeq 8$, with $\Delta z = 1$ (\textit{filled circles}, see text for details), and the comoving volume of an equivalent $\sim 1$~deg$^2$ survey with the Roman Space Telescope (RST, \textit{filled triangles}); the average number of PISNe found in our six \texttt{dustyGadget} cubes is also indicated in the plot (DG, \textit{empty squares}). A \textit{horizontal, dashed line} further indicates the reference value of 1 PISN per volume. The \textit{black line} refers to the average number of PISNe found in all haloes with $7.5 < \mathrm{Log}(M_\star/\si{M_\odot}) \leq 9.5$, while the \textit{coloured lines} refer to the average number found in haloes of different stellar mass bins (see Figure~\ref{fig:PISN}). \textit{Solid lines} refer to our reference model for Pop\ III stars, i.e. a Salpeter-like IMF in the range [100, 500]~\si{M_\odot} with no mass loss, while \textit{dotted lines} refer to the corresponding model assuming strong mass loss; the \textit{shaded area} shows the expected scatter for different assumed IMFs (see Figure~\ref{fig:popIIIIMFs} and Table~\ref{tab:popIIIIMFs}). As a reference, the \textit{purple pentagon} indicates the cumulative number of PISNe at $z > 8$ expected in a survey of $\sim$1~deg$^2$ in one year, computed by integrating over the observed PISN rate as a function of redshift from \citet[see their figure~15]{Jaacks_2019} in this redshift range, and re-normalizing by the considered survey area and by the extended IMF (see text and Table~\ref{tab:popIIIIMFs}).}
    \label{fig:PISN_vol}
\end{figure}

Figure~\ref{fig:PISN_vol} shows the average number of PISNe at $z = 8.1$ that we expect as a function of effective survey volume in our reference model, normalised by the efficiency $\eta_\mathrm{III}$; the expected scatter due to different choices of the IMF is also shown as a grey, shaded area. We multiply the values of $\overline{n}_\mathrm{PISN}$ in our simulations by the comoving volume of selected JWST surveys at $z \simeq 8$, with $\Delta z = 1$:
\begin{enumerate}
    \item the Next Generation Deep Extragalactic Exploratory Public (NGDEEP) Survey \citep{Finkelstein_2021, Pirzkal_2023, Bagley_2023};
    \item the Grism Lens-Amplified Survey from Space (GLASS\footnote{\url{https://glass.astro.ucla.edu/ers/}}, \citealt{Treu_2017, Treu_2022, Castellano_2022});
    \item the Ultradeep NIRSpec and NIRCam ObserVations before the Epoch of Reionization (UNCOVER\footnote{\url{https://jwst-uncover.github.io/}}, \citealt{Bezanson_2022, Furtak_2023, Weaver_2023});
    \item the Cosmic Evolution Early Release Science Survey (CEERS\footnote{\url{https://ceers.github.io/ceers-first-images-release}}, \citealt{Finkelstein_2017, Finkelstein_2022, Finkelstein_2023});
    \item the Public Release IMaging for Extragalactic Research (PRIMER\footnote{\url{https://primer-jwst.github.io/}}) survey \citep{Dunlop_2021};
    \item the PANORAMIC survey \citep{Williams_2021};
    \item the Cosmic Evolution Survey (COSMOS-Web\footnote{\url{https://cosmos.astro.caltech.edu/}}, \citealt{Casey_2023}).
\end{enumerate}
In our reference model, we expect less than 1 PISN on average in all the examined JWST surveys, even when considering all fields together. We also consider a possible $\sim 1$~deg$^2$ survey with the Wide Field Instrument (WFI) of the Nancy Grace Roman Space Telescope\footnote{\url{https://roman.gsfc.nasa.gov/}} in the same redshift range: we see that the expected number of PISNe in this volume in our reference model is $\simeq 1.5 \, \eta_\mathrm{III}$. The most pessimistic estimate (i.e. assuming a Salpeter-like IMF in the range [1, 1000]~\si{M_\odot}, see Table~\ref{tab:popIIIIMFs}) yields $0.1 \, \eta_\mathrm{III}$ PISNe, while this number can grow up to $\sim 2.3 \, \eta_\mathrm{III}$ in the most optimistic case\footnote{When the original IMFs in the range [1, 150]~\si{M_\odot} from \citet{Jaacks_2018_metalEnrichment} are considered, the upper limit on the total number of PISNe can grow up to $\sim 7.6 \, \eta_\mathrm{III}$.} (i.e. assuming a power-law IMF with $\alpha = -0.17$ in the range [1, 260]~\si{M_\odot}). Even higher survey volumes with Roman have been proposed in the literature: \citet{Moriya_2022_RST}, for example, suggested a 10~deg$^2$ transient survey with a limiting magnitude of 27.0~mag and 26.5~mag in the F158 and F213 ﬁlters respectively, to be conducted for ﬁve years with a cadence of one year with the aim of looking for PISNe candidates during the EoR (also see Section~\ref{sec:results_PISNeVSStars_integratedEmission}). This would further increase the expected number of PISNe by a factor ten.

We emphasize that in this paper we investigate the possibility of detecting PISNe during the EoR and using them as a marker for Pop\ III stars. Hence, we focus on galaxies that are potentially observable with JWST and Roman, so that we can also study the underlying stellar populations in details. However, mini-haloes are the most favourable environments for Pop\ III star formation at high redshifts, meaning we expect a higher number of PISNe occurring in low-mass halos closer to Cosmic Dawn (also refer to the first caveat of Section~\ref{sec:discussion_caveats}). We can compute the cumulative observed PISN rate per unit time per unit solid angle at $z > 8$ as:
\begin{equation}
    \frac {\dd \overline{N}_\mathrm{PISN}}{\dd t \dd \Omega} (z > 8) \simeq \frac{\overline{N}_\mathrm{PISN}}{M_\mathrm{III}}  \int_{z = 8}^{\infty} \frac{\Psi_\mathrm{III}(z)}{1 + z} r^2(z) \diff{r}{z} \dd z,
    \label{eq:PISNeFromSFRD}
\end{equation}
where $\Psi_\mathrm{III} (t)$ is the Pop\ III cosmic star formation rate density (SFRD) at redshift $z$, and $r(z)$ is the comoving distance to redshift $z$ (see e.g. \citealt{Hummel_2012, Jaacks_2019}). In Figure~\ref{fig:PISN_vol}, we show the total number of PISNe up to Cosmic Dawn that would be found in 1~yr in a survey area of $\sim$1~deg$^2$ (purple pentagon), computed by integrating over the observed PISN rate as a function of redshift shown in figure~15 of \citet{Jaacks_2019}\footnote{Their model is able to better resolve star formation in low-mass halos at high redshifts ($z \gtrsim 13$), that we tend to underestimate systematically due to our coarse resolution. See e.g. figure~A1 of \citealt{Venditti_2023} for a comparison with the Pop\ III SFRD of various high-resolution models and simulations on smaller boxes (box side $\lesssim 4h^{-1}$~cMpc).}. An approximate number of $\simeq 26$ PISNe/yr can be found on average at $z > 8$ from their model\footnote{Note that the total mass of Pop\ III stellar populations in this simulation is $\sim 1000 ~ \si{M_\odot}$, much smaller than our stellar populations of $\sim 10^6 ~ \si{M_\odot}$. Hence, this number should not be rescaled by the efficiency $\eta_\mathrm{III}$ (i.e. a value of $\eta_\mathrm{III} = 1$ should be considered). We also re-normalize by our considered survey area.}, reduced to $\simeq 10$ when accounting for a more extended IMF up to $\simeq 260 ~ \si{M_\odot}$, the inferred mass of the PISN progenitor found in \citet{Xing_2023} (mod-Larson IMF in Table~\ref{tab:popIIIIMFs}).

\subsection{Detecting PISNe and discriminating against stellar emission}
\label{sec:results_PISNeVSStars}

\subsubsection{Integrated emission}
\label{sec:results_PISNeVSStars_integratedEmission}

To estimate the chance of observing PISNe against the stellar emission of their hosting galaxies, in Figure~\ref{fig:PISN_Lbol} we compare the predicted bolometric light curves of representative PISNe with the bolometric luminosity arising from the stellar populations in galaxies of given stellar mass. 

\begin{figure}
    \centering
    \includegraphics[width=\linewidth]{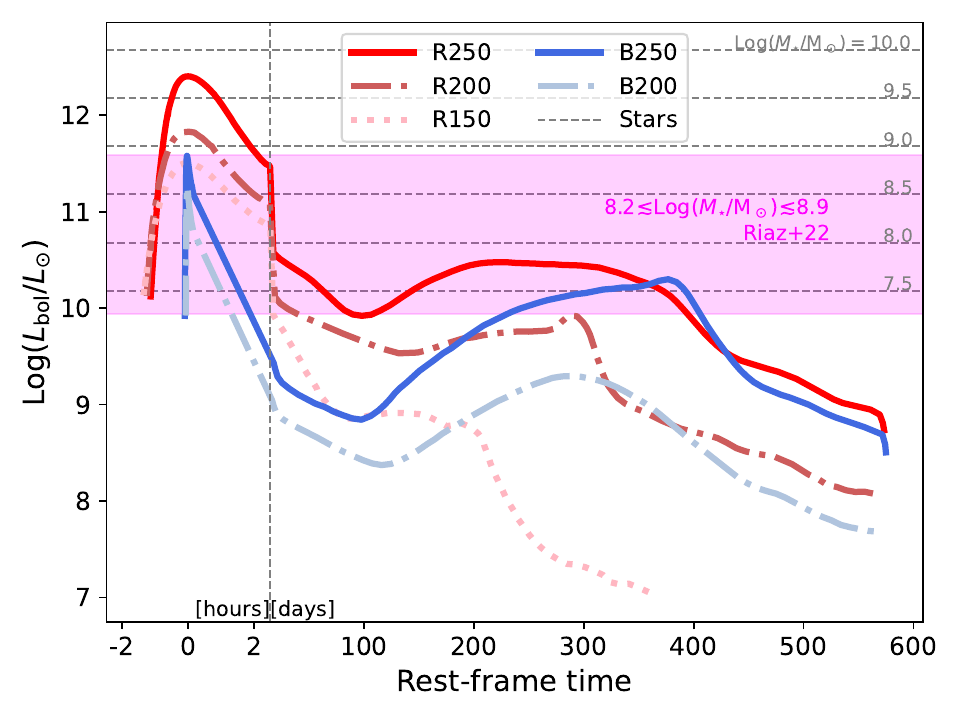}
    \caption{Total bolometric luminosity $L_\mathrm{bol}$ of galaxies with various stellar masses $M_\star$ (\textit{horizontal, dashed, light-grey lines}), compared with the light curve of PISNe with different RSG/BSG progenitors as a function of rest-frame time (see text for details); the left-hand side of the plot (before the \textit{vertical, dashed line}) shows the brief, ``shock-breakout'' phase (from figure~3 of \citealt{Kasen_2011}), expanded to display the time evolution in units of hours, while the right-hand side shows the long-term emission in day units (from figure~5 of \citealt{Kasen_2011}). The $L_\mathrm{bol}$ of galaxies with given $M_\star$ is computed assuming all the stellar mass consists of Pop\ II/I stellar populations following a Salpeter IMF in the range [0.1, 100]~\si{M_\odot}, and that all stars obey standard mass-luminosity relations (Equation~\ref{eq:mass-luminosity_populations}, valid for stars at solar metallicity on the ZAMS, with $C = 10^{2.68} ~ \si{L_\odot.M_\odot^{-1}}$). The \textit{magenta, shaded region} also shows a comparison with the average stellar luminosity of haloes with virial mass $\sim 10^{11} ~ \si{M_\odot}$ at $6 \lesssim z \lesssim 30$, corresponding to $M_\star \sim 10^{8.2} - 10^{8.9} ~ \si{M_\odot}$, from \citet[see their figures~1 and~2]{Riaz_2022}.}
    \label{fig:PISN_Lbol}
\end{figure}

We consider PISNe with five different progenitor stars from \citet{Kasen_2011}, particularly two zero-metallicity stars dying as compact blue super-giants (BSGs, i.e. B250 and B200, respectively with initial masses of 250~\si{M_\odot} and 200~\si{M_\odot}) and three $10^{-4} ~ \si{Z_\odot}$ stars dying as red super-giants\footnote{The authors argued that the difference between the two expected outcomes also relies on the choice of other uncertain parameters such as primary nitrogen production and mixing, and that with typical values most zero-metallicity stars would also end their life as RSGs.} (RSGs, i.e. R250, R200 and R150, with 250~\si{M_\odot}, 200~\si{M_\odot} and 150~\si{M_\odot} respectively), with radii 10-50 times larger. The highest luminosities are reached during the ``shock-breakout'' phase of the explosion, when the radiation-dominated shock, resulting from the expansion of the exploded He core into the hydrogen envelope, approaches the surface of the star and escapes in a luminous X-ray/UV burst. The peak luminosity for each SN can reach values higher than $10^{12} ~ \si{L_\odot}$ (see table~2 of \citealt{Kasen_2011}). However, this phase only lasts a few hours. We are instead more interested in the longer-lasting emission following the breakout, as radiation continues to diffuse out of the expanding, cooling ejecta and it may be visible at rest-frame optical wavelengths for several weeks, and even longer in the IR (see e.g. figure~7 of \citealt{Kasen_2011}). The initial phase of the light-curve is powered by the diffusion of thermal energy deposited by the shock and it is dimmer for the BSG models due to the relatively small radii of the progenitors; the secondary peak results from the radioactive decay of synthesized $^{56}$Ni.

To roughly estimate the luminosity arising from the stellar populations in galaxies of given stellar mass, we assume the luminosity of a given star of mass $m$ in the range $[m_i, ~ m_{i+1}]$ can be inferred from a simple mass-luminosity relation of the form
\begin{equation}
    L^{(i)}(m) = a_i m^{b_i}
    \label{eq:mass-luminosity_stars},
\end{equation}
\citep[see, e.g.][]{Riaz_2022}\footnote{The spread between the average stellar luminosities of the most massive haloes in \citet{Riaz_2022} is also shown as a reference in Figure~\ref{fig:PISN_Lbol}.}. The luminosity arising from all stars within this mass range in a given stellar population can then be obtained by integrating over our assumed IMF:
\begin{equation}
    L_\star^{(i)} = M_\star \frac{\int_{m_i}^{m_{i+1}} L^{(i)}(m) \phi(m) \dd m}{\int_{m_\mathrm{low}}^{m_\mathrm{up}} m \phi(m) \dd m}.
\end{equation}
Hence, we find a simple relation between the total stellar mass of the haloes $M_\star$ and their luminosity $L_\star$:
\begin{equation}
    L_\star = \sum_i L_\star^{(i)} =  C \times M_\star,
    \label{eq:mass-luminosity_populations}
\end{equation}
with $C = \left[ \sum_i \int_{m_i}^{m_{i+1}} L^{(i)}(m) \phi(m) \dd m \right] / \left[ \int_{m_\mathrm{low}}^{m_\mathrm{up}} m \phi(m) \dd m \right]$. Since we are interested in the total emission, as a first approximation we assume all the stellar mass consists of Pop\ II/I stars and neglect the contribution of Pop\ III\footnote{Note that this is a reasonable approximation only when considering the total bolometric emission, while the contribution of massive Pop\ IIIs to the detailed spectral energy distribution, especially in the ionising bands, may be non-negligible \citep{Tumlinson_Shull_2000, Tumlinson_2001, Bromm_2001_spectra, Schaerer_2002, Schaerer_2003, Raiter_2010}.}, making up less than 5\% of the total stellar mass (see e.g. figure~3 of \citealt{Venditti_2023} and the lower curves of figure~3 in \citealt{Riaz_2022}). We use standard mass-luminosity relations\footnote{The average mass-luminosity relation for stars with masses $1\,\si{M_\odot} \lesssim m \lesssim 50\,\si{M_\odot}$ is taken from \citet[p. 253]{Kippenhahn_2013}. The relation at low stellar masses, for which convection is the dominant energy transport process, is taken from \citet[p. 19]{Duric_2003}, as inferred from observations of stars in the Hyades cluster \citep{Torres_1997}. At high masses, as the stars approach the Eddington limit and become unstable, quickly losing mass by intense stellar winds due to the increasing radiation pressure \citep{Kudritzki_Puls_2000}, the relation becomes progressively flatter, approximated as linear in the limit $m \gtrsim 50~\si{M_\odot}$.} for Pop\ II/I stars (in Eq.~\ref{eq:mass-luminosity_stars}):
\begin{equation}
    L(m) \simeq 
    \begin{cases}
    L_{50} (m/50~\si{M_\odot}) & m \gtrsim 50~\si{M_\odot}, \\
    \si{L_\odot} (m/\si{M_\odot})^{3.35} & 1~\si{M_\odot} \lesssim m \lesssim 50~\si{M_\odot}, \\
    \si{L_\odot} (m/\si{M_\odot})^{4.0} & 0.4~\si{M_\odot} \lesssim m \lesssim 1~\si{M_\odot}, \\
    L_{0.4} (m/0.4~\si{M_\odot})^{2.3} & m \lesssim 0.4~\si{M_\odot},
    \end{cases}
    \label{eq:mass-luminosity_PopII}
\end{equation}
where $L_{50}$ and $L_{0.4}$ are, respectively, the luminosity of a star with 50~\si{M_\odot} and 0.4~\si{M_\odot}. By integrating these relations over our assumed Salpeter IMF in the range [0.1, 100]~\si{M_\odot}, we find $C = 10^{2.68} ~ \si{L_\odot.M_\odot^{-1}}$ (Equation~\ref{eq:mass-luminosity_populations}).

\begin{figure*}
    \centering
    \includegraphics[width=\linewidth]{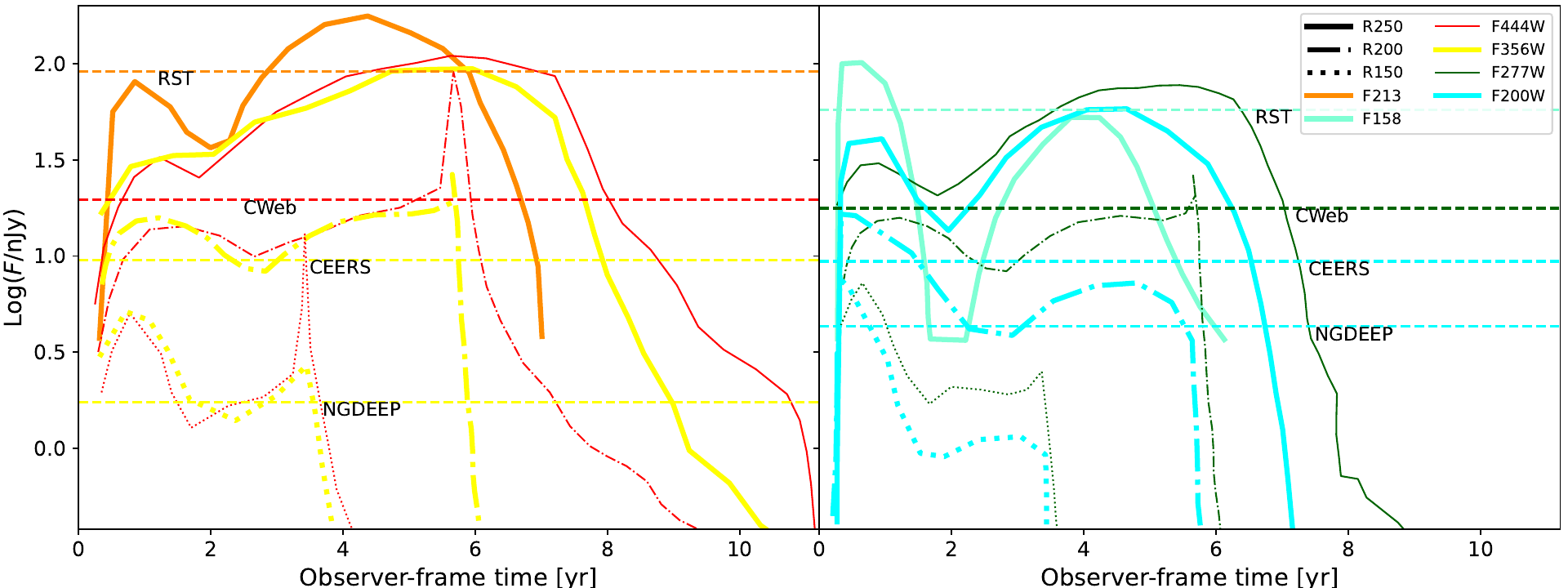}
    \caption{Expected observed light curve of PISNe from the different RSG progenitors of Figure~\ref{fig:PISN_Lbol} in the same F356W and F200W filters (\textit{thick, yellow lines} and \textit{thick, cyan lines}), in the F444W and F277W filters (\textit{thin, red lines} and \textit{thin, dark-green lines}), and in the F213 and F158 filters of the Roman Space Telescope (\textit{thick, orange lines} and \textit{thick, aquamarine lines}), at $z \simeq 6$, as a function of observer-frame time (from figure~4 of \citealt{Hartwig_2018} and figure~6 of \citealt{Moriya_2022_RST}). The sensitivity limits of selected JWST surveys and the limiting sensitivity necessary to separate SNeIa and SNeII at $z > 1.5$ from PISNe and SLSNe at $z > 6$ with Roman (see text for details) are indicated as \textit{horizontal, dashed lines}, with the same colour legend used for the filters.}
    \label{fig:PISN_JWST+RomanFilters}
\end{figure*}

Note that the relations in Equation~\ref{eq:mass-luminosity_PopII} are only valid for the bolometric luminosities of stars on the Zero Age Main Sequence (ZAMS), while older stars would be found at progressively lower luminosities. Therefore, these results should be strictly interpreted as upper limits. Keeping this caveat in mind, we see from Figure~\ref{fig:PISN_Lbol} that the peak emission from most PISNe would easily outshine the stellar emission of their hosting haloes, at least for $M_\star \lesssim 10^9 ~ \si{M_\odot}$, especially when considering the most likely RSG-scenarios. However, as previously emphasised, this emission is short-lived with respect to the total light curve, and our chance of catching the event close to its peak is of the order of $\sim 1$~hour/1~yr$\sim 10^{-4}$. Considering the long-term evolution, when the luminosity can drop by more than two orders of magnitude, we have a much less favourable scenario, with only the extreme R250 PISN clearly rising above the stellar emission from $M_\star \lesssim 10^8 ~ \si{M_\odot}$ haloes.

Figure~\ref{fig:PISN_JWST+RomanFilters} shows the observed PISN light curves for the RSG-progenitor models at $z \simeq 6$, in four JWST/NIRCam filters (F444W, F356W, F277W and F200W), as well as in two filters of the Nancy Grace Roman Space Telescope (F213 and F158). The combination of F356W and F200W was suggested by \citet{Hartwig_2018} as the optimal 2-filter diagnostic to detect the prompt emission of PISNe at $6 \lesssim z \lesssim 12$ with JWST, thus identifying possible candidates to be confirmed as transients through follow-up observations. This choice of filters maximizes the visibility time (i.e. the fraction of the light curve that we are able to observe) and thus the probability of finding a PISN, with an optimal exposure time of $\sim 600$~s (see figure~6 of \citealt{Hartwig_2018}). Similarly, \citet{Moriya_2022_RST} determined that F158-F213 is the best two-filter combination to discriminate between the classical SNeIa/SNeII at $z > 1.5$ and PISNe/SLSNe at $z > 6$ with Roman, by comparing the color-magnitude diagrams of SNe at all phases. 

The sensitivity limits\footnote{The sensitivity limits are from (i) table~1 of \citet{Casey_2023} for COSMOS-Web; (ii) \url{https://ceers.github.io/obs.html} for CEERS; (iii) table~1 of \citet{Bagley_2023} for NGDEEP.} in the reference filters of CEERS and of the shallowest/deepest JWST surveys (COSMOS-Web/NGDEEP) included in Figure~\ref{fig:PISN_vol} are indicated as horizontal, dashed lines in the plot; for COSMOS-Web, as the optimal filters F356W and F200W are not available, we report the limits of the closest filters in terms of wavelength coverage, i.e. F444W and F277W\footnote{Note however that these filters cannot be exposed simultaneously. The only two-filter diagnostics available for COSMOS-Web would be F150W-F277W and F150W-F444W, but these would be among the least efficient combinations (see figure~6 of \citealt{Hartwig_2018}).}. We see that the PISNe arising from the most massive progenitors would be visible from a time $\lesssim 2$~months (in the case of R200 in the COSMOS-Web/F277W filter) up to a time $\sim 9$~yr (in the case of R250 in the NGDEEP/F356W filter). The least-massive-progenitor PISN (R150) would also be visible for almost one year in NGDEEP/F200W and almost two years in NGDEEP/F356W. The limiting sensitivity necessary to separate SNeIa and SNeII at $z > 1.5$ from PISNe and SLSNe at $z > 6$ with Roman (see \citealt{Moriya_2022_RST}) is also indicated: with these sensitivities, the most massive PISN would be observable for about three years in F213 and for about a year and a half in F158.

\subsubsection{Resolved observations}
\label{sec:results_PISNeVSStars_resolvedObservations}

\begin{figure}
    \centering
    \includegraphics[width=\linewidth]{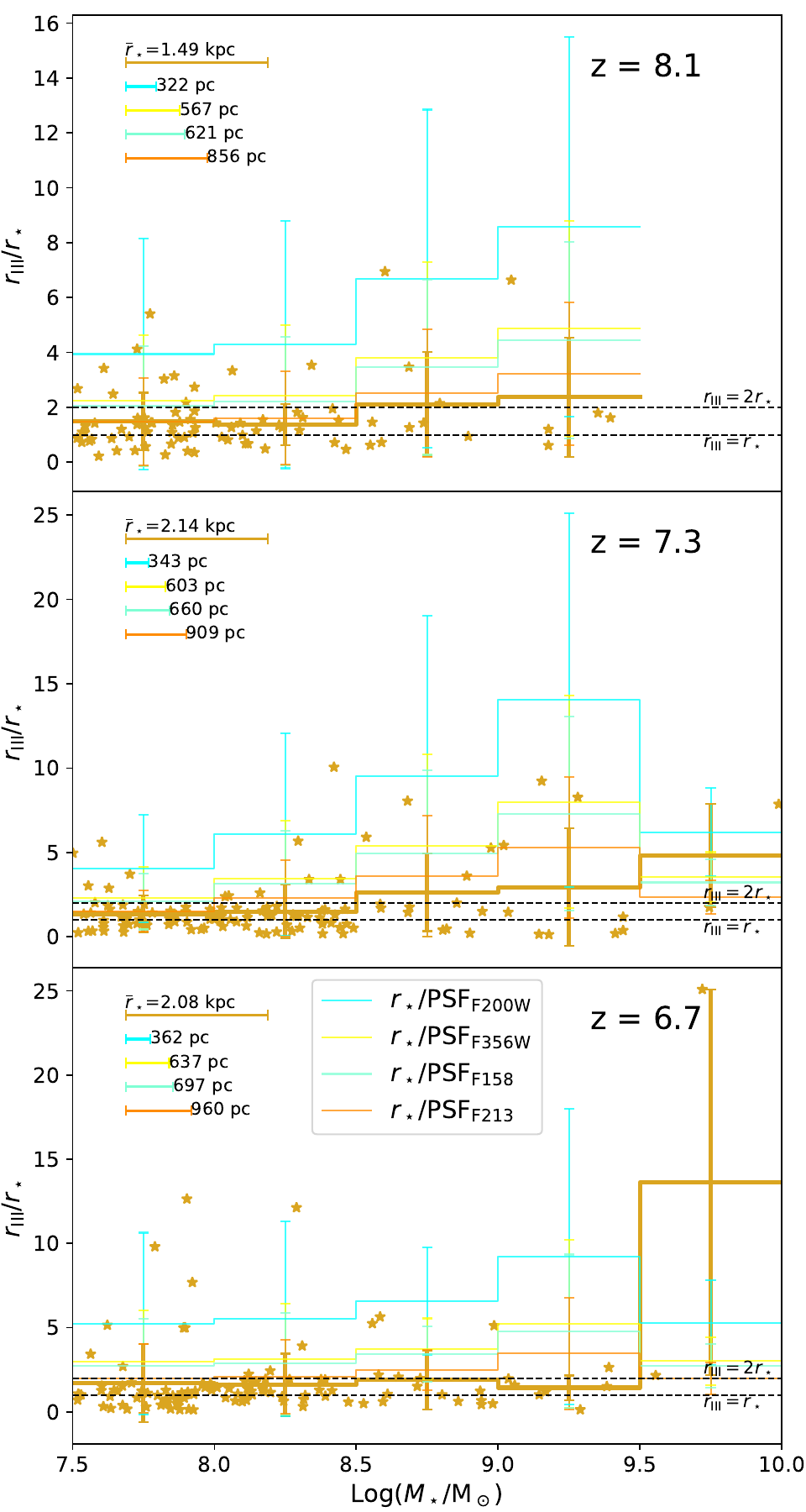}
    \caption{Ratio between the Pop\ III distance from the centre of mass of the hosting halo $r_\mathrm{III}$ and the stellar mass-weighted radius $r_\star$ of the halo, for each Pop\ III stellar particle that could potentially produce PISNe (\textit{gold stars}, i.e. Pop\ III particles with age $\tau_\mathrm{noML} (260~\si{M_\odot}) \leq t_\mathrm{III} \leq \tau_\mathrm{strongML} (140~\si{M_\odot})$). The mean and standard deviation computed in five bins of $M_\star$ (with a spacing of 0.5~dex in the range $7.5 \leq \mathrm{Log} M_\star / \si{M_\odot} < 10$) is also shown as a \textit{gold, thick, solid line}. Results are reported at $z = 8.1$ (\textbf{top panel}), $z = 7.3$ (\textbf{middle panel}) and $z = 6.7$ (\textbf{bottom panel}). The \textit{black, dashed lines} indicate where $r_\mathrm{III} = r_\star$ or $r_\mathrm{III} = 2 r_\star$; Pop\ IIIs with $r_\mathrm{III}$ above these thresholds (depending on the strictness of the adopted criterion) are considered ``external''. The mean and standard deviation of $r_\star$ normalized to the PSF of selected JWST/NIRCam and Roman/WFI filters (F200W/F356W and F158/F213 respectively) are also shown in the plots as \textit{thin, coloured lines}, with same colour legend of Figure~\ref{fig:PISN_JWST+RomanFilters}. The segments at the top-left corners indicate the relative size of the filters PSF (\textit{coloured segments}) compared to the average value of $r_\star$ over the considered sample (\textit{gold segments}); the absolute size is also indicated next to each segment.}
    \label{fig:PISN_ext}
\end{figure}

In Figure~\ref{fig:PISN_Lbol} we considered the integrated emission arising from all stellar populations in the galaxies, but in reality it is unlikely that this emission would be concentrated in a small region near their centre. In \citet{Venditti_2023} we discussed the complex and irregular morphology of early galaxies in \texttt{dustyGadget} simulations in terms of their gas, dust and stellar components, as they are mostly still in an assembly phase. Moreover, given the low metallicity required for their formation, many Pop\ IIIs will be found either at the periphery of Pop\ II clusters or in pristine satellites of their hosting haloes. This could result in lower levels of contamination from higher-metallicity stellar populations to the signal of Pop\ IIIs and of their PISNe, provided that we are able to observe these galaxies with enough spatial resolution.

In Figure~\ref{fig:PISN_ext} we show the ratio between the distance\footnote{As a single stellar particle represents a cluster of Pop III stars born together surrounded by gas, and we cannot follow their dynamics within this resolution element of $\sim$\,few pc, the estimated error on $\sim$\,kpc scales is expected to be of order $10^{-3}$.} of Pop\ III particles\footnote{Note that in this figure we consider the largest sample of potential candidates available, i.e. Pop\ III particles with age $\tau_\mathrm{noML} (260~\si{M_\odot}) \leq t_\mathrm{III} \leq \tau_\mathrm{strongML} (140~\si{M_\odot})$, with $\tau_\mathrm{noML} (260~\si{M_\odot}) = 2.10$~Myr the lifetime of a $260~\si{M_\odot}$ Pop\ III star assuming no mass loss and $\tau_\mathrm{strongML} (140~\si{M_\odot}) = 3.06$~Myr the lifetime of a $140~\si{M_\odot}$ Pop\ III star assuming strong mass loss. Further note that the displayed quantities are independent of the chosen IMF and $\eta_\mathrm{III}$.} that may produce PISNe at a given redshift ($z = 8.1$, 7.3 and 6.7) from the centre of mass of their hosting halo and the stellar-mass-weighted radius $r_\star$, as a function of the halo stellar mass $M_\star$. We see that there is a large scatter in the values, and that Pop\ IIIs can be found at a distance up to twenty-five times $r_\star$, in the most extreme case. To highlight the trend of $r_\mathrm{III} / r_\star$ as a function of $M_\star$, we also plot the mean and standard deviation in the same $M_\star$ bins of Figure~\ref{fig:PISN}. Most particles have an $r_\mathrm{III}$ slightly higher than $r_\star$, although if we consider a stricter criterion to indicate ``external'' PISNe (i.e. $r_\mathrm{III} > 2 r_\star$), only particles in haloes higher than $10^{8.5} ~ \si{M_\odot}$ satisfy this requirement on average, and only particles in haloes higher than $10^{9.5} ~ \si{M_\odot}$ at $z = 6.7$. Note, however, that low-mass galaxies also have a generally lower luminosity, and therefore they are less likely to outshine their hosted PISN (see Figure~\ref{fig:PISN_Lbol}). Despite the large scatter, there is indeed a slightly growing trend of $r_\mathrm{III} / r_\star$ with increasing $M_\star$. 

We compare the physical size of our galaxies to the resolution element of the selected JWST/NIRCam and Roman/WFI filters discussed in Section~\ref{sec:results_PISNeVSStars_integratedEmission}. The physical size of the Point Spread Function (PSF) at given redshift is $s(z) = \theta D_\mathrm{A}(z)$, with $D_\mathrm{A}(z)$ the angular diameter distance at redshift $z$ and $\theta$ the angular size\footnote{The PSF of NIRCam/F200W and NIRCam/F356W is taken from \url{https://jwst-docs.stsci.edu/jwst-near-infrared-camera/nircam-performance/nircam-point-spread-functions}, while the PSF of WFI/F158 and WFI/F213 is taken from \url{https://roman.gsfc.nasa.gov/science/WFI_technical.html}.} of the PSF. In Figure~\ref{fig:PISN_ext}, we show the mean and standard deviation of $r_\star$ normalized to the PSF in the same bins of $M_\star$. For reference, we also show the size of the various PSFs relative to the average $r_\star$ of our PISN-hosting galaxies at each redshift. We see that, on average, the bulk of our galaxies is resolved with more than one pixel in all the considered filters. Particularly, they are best resolved in NIRCam/F200W, while WFI/F213 has the coarsest spatial resolution. High-mass galaxies up to $10^{9.5} ~ \si{M_\odot}$ are also generally better resolved than their lower-mass counterparts.

\section{Discussion}
\label{sec:discussion}

\subsection{Main caveats}
\label{sec:discussion_caveats}

The following caveats should be kept in mind:

\begin{enumerate}
    \item the statistics for high-mass, well-resolved haloes in our simulations is quite scarce. More reliable results for the highest-stellar-mass bins in Figure~\ref{fig:PISN} would require even larger simulated volumes. We also did not take into account the contribution of galaxies with $M_\star < 10^{7.5} ~ \si{M_\odot}$, that are not well resolved in our simulations due to our mass resolution limits, and that would also be the most affected by radiative feedback (see point (v)). Neglecting this halo population, that hosts a significant portion of the total Pop\ III mass ($\gtrsim 60 \%$ at the considered redshifts, see \citealt{Venditti_2023}), will result in an underestimation of the total number of expected PISNe, at least by a factor 1.5. However, we remark that we are mainly interested in PISN hosts in which we can potentially study the underlying stellar populations in details, to look for candidate Pop\ III stars during the EoR. A full picture on all the PISNe produced in our cubes regardless of their environment is beyond the scopes of the present work.
    \item In Equation~\ref{eq:avNumberOfPISNe_atFixedTime}, we are assuming for simplicity that PISNe have uniform probability of occurring within $\Delta t_\mathrm{PISN}$ for a given stellar population. However, in reality this probability depends on the assumed IMF; in particular, with our IMF, lower-mass Pop\ III stars are more likely to form, therefore we should expect more PISNe arising from stars closer to the lower-mass threshold for PISN progenitors, and hence having longer lifetimes. Moreover, a random sampling of the IMF would be required for a more realistic estimate of the average number of PISNe produced per stellar population.
    \item We have explored two different models with different assumptions for the stellar mass loss arising from Pop\ IIIs, however, many other uncertainties exist. For example, we have not explored different choices for the value of $Z_\mathrm{crit}$, although the chosen assumptions may change the results for Pop\ III star formation at cosmological scales considerably (see e.g. \citealt{Maio_2010}). We did consider the impact of choosing different IMFs in the estimate of the average number of PISNe produced per unit Pop\ III mass (Equation~\ref{eq:avNumberOfPISNePerUnitMass}). Nonetheless, we emphasize that a full exploration of the impact of IMF variations would require a self-consistent approach.
    \item Our chosen Pop\ III IMF covers the mass range [100,~500]~\si{M_\odot}. On the other hand, studies on the properties of carbon-enhanced extremely metal-poor stars show that they are best matched by the nucleosynthetic yield of traditional core-collapse SNe\footnote{However, some level of contamination of enrichment from Pop\ II stars is also possible \citep{Ji_2015, Vanni_2023}.} \citep{2Iwamoto_2005, Keller_2014, Ishigaki_2014, deBennassuti_2014, deBennassuti_2017, Fraser_2017, Magg_2022, Aguado_2023_mostMetalPoorStar}, suggesting that Pop\ III progenitors with masses $\sim 10 - 40 ~ M_\odot$ are indeed more prevalent. An even lower minimum value of $\sim 1 ~ \si{M_\odot}$ has been suggested by several works of stellar archaeology at a high confidence level \citep{Hartwig_2015, Rossi_2021}. Here, we are not interested in the detailed nucleosynthetic pattern, hence we do not require a high precision on the IMF as in classical archaeology studies and we do not expect our conclusions to be significantly affected by this assumption. Indeed, the main impact of the IMF on our results is through the parameter $\overline{N}_\mathrm{PISN}/M_\mathrm{III}$, and the value of $2.17 \times 10^{-3} ~ \si{M_\odot}$ that we find with our assumed IMF is within the allowed range spanned by IMFs extended down to $\sim 1~\si{M_\odot}$. We also note that, while considering an IMF more biased towards massive stars might cause in principle an overproduction of PISNe, the reality is more complex. In fact, given the exceptionally high metal yield of PISNe, this choice may actually lead in the opposite direction, as a faster enrichment might induce an earlier transition to the Pop\ II mode of star formation. The effect of the Pop\ III IMF on cosmic star formation history is debated \citep{Maio_2010, Maio_2016, Pallottini_2014}. In \citet{Venditti_2023}, we found that our Pop\ III SFRD is in line with other models and simulations, although large uncertainties are involved.
    \item Even though the simulation includes an homogeneous UV background as in \citet{Haardt_Madau_1996}, this does not have feedback on cosmic star formation at $z > 6$, i.e. in the epoch considered in the present work. The lack of a proper treatment of radiative feedback, - especially in the UV and LW bands - certainly affects the late Pop III star formation. This limitation is extensively discussed in \citet{Venditti_2023}, in which our Pop III star formation rate density is also carefully compared with the results of other small-scale simulations modelling the local/global effects of UV and LW radiation.
    \item At the current stage, \texttt{dustyGadget} does not include a model for metal mixing and turbulent metal diffusion below our gas mass resolution. While allowing for incomplete inhomogeneous mixing at the sub-grid level may allow to enhance Pop~III star formation by a factor 2-3 \citep{Sarmento_2016, Sarmento_2017, Sarmento_2018, Sarmento_Scannapieco_2022}, the study by \cite{Su_2017} showed that sub-grid turbulent metal diffusion has a negligible impact on general star formation and ISM properties. The effect of diffusion has also been shown to be weaker at lower metallicity and relatively unimportant for Pop\ III star formation \citep{Jeon_2017}. We refer to \citet{Venditti_2023} for further discussion and for a comparison with the models of \citet{Sarmento_2018} and \citet{Sarmento_Scannapieco_2022}.
    \item We have not discussed our ability to discriminate PISNe from other theoretical red sources such as dark stars \citep{Spoylar_2008, Freese_2008, Freese_2010, Yoon_2008, Natarajan_2009, Hirano_2011, Banik_2019} and direct-collapse black holes \citep{Hammerle_2020}. We note, however, that we should be able to distinguish these sources through follow-up observations after a few years, due to the time variability of PISNe.
\end{enumerate}

\subsection{Implications of future detections}
\label{sec:discussion_futureDetections}

In light of our results, we wish to discuss the implications of future detections/non-detections of PISNe with both JWST and Roman. In our model, we predict that we should be able to see some PISNe at high-redshift at least through wide surveys with the Roman Space Telescope. Hence, the following questions arise: what if we do not find any PISN at high redshift? What if, conversely, we find a surprisingly high number of PISNe?

A higher number of PISNe than allowed by our models may be explained if a substantial fraction of Pop\ II stars is also able to reach sufficiently high masses. The JWST capabilities for discovering PISNe from 0.14–0.43~\si{Z_\odot} stars at $z < 10$ have been discussed e.g. in \citet{Regos_2020}. Although the general consensus is that Pop\ II/I stars usually follow a standard Salpeter IMF, with masses $\sim 0.1 - 100 ~\si{M_\odot}$, our constraints on the IMF are mainly derived from observations of the local Universe. On the other hand, more top-heavy IMFs might actually be more common at high redshifts (see e.g. \citealt{Chon_2022}). The prevalence of a top-heavy IMF has been suggested to alleviate the tension between JWST observations and existing models and simulations, which seem to systematically under-produce massive, bright galaxies at high redshifts ($z \gtrsim 10$, \citealt{Yung_2023, Harikane_2023, Trinca_2023}). \citet{Olivier_2022} also demonstrated that a high-temperature black-body spectrum ($T_\mathrm{eff} \sim 80000$~K, associated to very massive stars) is necessary to explain the ionization level of low-$z$, extreme-emission-line galaxies that are considered analogues to EoR galaxies.
We further note that we focused on our chance of observing PISNe through their prompt emission over a time $\Delta t_\mathrm{prompt} \sim$~1~yr, hence neglecting the possibility that PISNe may be observable through their afterglow on much longer timescales. \citet{Hartwig_2018} predict that the PISN afterglow is not accessible through current observational facilities, although observatories and telescopes with stronger capabilities may be able to detect it in the future. However, more accurate modelling may find that this signal is actually brighter than these predictions, implying that the $\Delta t_\mathrm{prompt} \longrightarrow \Delta t_\mathrm{afterglow}$ in Equation~\ref{eq:avNumberOfPISNe_atFixedTime} - and hence the expected number of PISNe - can be boosted up to a hundred times.

Not finding any PISNe during the EoR with either JWST or Roman would also be surprising, and it would require a revision of our models for Pop\ III stars. Maybe Pop\ IIIs do not reach high enough masses to produce a significant number of PISNe, either because the physics of star-forming clouds only allows stellar masses lower than $\sim 140 ~ \si{M_\odot}$, or because the stars lose most of their mass before dying. An alternative explanation is that Pop\ III star formation is suppressed much earlier than $z \sim 10$. As most models and simulation currently predict that the inhomogeneous metal enrichment should allow a late Pop\ III star formation, our models for chemical and radiative feedback may also need to be reconsidered in this scenario.

\subsection{Detection strategies: archaeology vs. direct detections at high $z$}
\label{sec:discussion_detectionStrategies}

In the introduction of this paper we listed many possible strategies to study PISNe and their Pop\ III progenitors. These include direct detection, especially at high redshift where massive Pop\ III stars may still be actively forming, and cosmic archaeology \citep{Aoki_2014, Salvadori_2007, Yoshii_2022, Aguado_2023_PISN-explorer, Xing_2023}. Both techniques present unique advantages and challenges.

The first obvious advantage of cosmic archaeology is that it makes the study of PISNe accessible through observations in the local Universe. However, the nucleosynthetic footprint of low-metallicity gas may result from a complex stratification of multiple sources of metal pollution, that can be hard to disentangle \citep[e.g.][]{Ji_2015, Vanni_2023}. Moreover, an inherent issue lies in the selection of candidate PISN descendants: given its significant metal yields, a single PISN may cause its environment to ``overshoot'' and promptly reach a relatively high metallicity of $\gtrsim 10^{-3} ~ \si{Z_\odot}$ \citep{Karlsson_2008, Greif_2010, Wise_2012_PMEnrichment}, and all PISN-descendant candidates to date \citep{Aoki_2014, Salvadori_2019, Xing_2023, Aguado_2023_PISN-explorer} exhibit in fact $\mathrm{[Fe/H]} > - 2.5$. This means we should not only look at the most metal-poor stars/environments, rendering the identification and interpretation of potential candidates even more difficult (see also \citealt{deBennassuti_2017}). We certainly need improved selection strategies \citep{Aguado_2023_PISN-explorer}, possibly relying on photometry rather than costly spectroscopy on each source.

Throughout this paper, we demonstrated that a direct detection of PISNe during the EoR is not beyond hope. The identification of PISNe in this case is probably more straightforward and it has been already discussed in the literature \citep{Kasen_2011, Hartwig_2018, Moriya_2022_RST}. We also note that no SN of any kind has yet been found at $z \gtrsim 3$ \citep[e.g.][]{Cooke_2012,Jones_2013}. Due to our sensitivity limits, it is actually more likely that if we do find SNe at high redshifts, these will be either PISNe or other kinds of SLSNe. The main obstacle is the fundamentally transient nature of SNe: although PISNe light curves can be stretched up to $\sim 10$~yr in the observer frame, this transient nature, combined with their rarity, make their detection less likely. A longer visibility time might be attained through their afterglow (see e.g. the discussion in Section~\ref{sec:discussion_futureDetections}). The direct imprints of SN feedback may also be studied as a fossil record for longer time scales after the explosion. If we consider for example a galaxy with $M_\star \sim 10^7 ~ \si{M_\odot}$, of which $\sim 10^4 ~ \si{M_\odot}$ in Pop\ IIIs, $\sim 40\%$ of the Pop\ III mass would end up as PISNe with our assumed IMF, while $\sim 10\%$ of the Pop\ II mass would end up as a traditional core-collapse SNe. By assuming a return fraction of $\sim 0.1$ for core-collapse progenitors and $\sim 0.5$ for PISN progenitors \citep{Karlsson_2013}, the ratio of metals from PISNe vs core-collapse would be of the order of $2 \times 10^{-2}$, i.e. non negligible. This means we may be able to reliably tell apart the chemical signature of PISNe outflows. A thorough investigation of such matters is left to future works.

Looking for PISNe at even higher redshifts ($z \gtrsim 10$) may also be feasible thanks to the efficient Pop\ III formation in low-mass, pristine halos towards Cosmic Dawn (see e.g. the discussion at the end of Section~\ref{sec:results_PISNNumber}). We did not investigate such possibility in depth as the galaxies hosting these PISNe are less likely to be observable, while in the current work we were mainly interested in using PISNe as tracers for Pop\ III stars. Nonetheless, the tentative detection of HeII emission in the vicinity of a luminous $z = 10.6$ galaxy \citep{Maiolino_2023} might already indicate the presence of observable Pop\ III clusters at these redshifts. Dedicated surveys aimed at finding PISNe at very high redshifts may reveal even more of these fascinating events.

In conclusion, many promising strategies to look for PISNe at high redshifts are now on the horizon. So far, stellar archaeology has been our best tool. While this technique will continue to provide invaluable information, ultimately, the combination of all the proposed diagnostics of PISNe will serve our purposes, as they can provide independent constraints on the rate of PISNe across time and on the properties of their Pop\ III progenitors. We stress that a synergistic approach, exploiting the different strengths of all these methods, will give us the best chance to gain a complete picture on massive Pop\ III stars.

\section{Summary and Conclusions}
\label{sec:conclusions}

In this paper we studied the probability of finding PISNe arising from Pop\ III stars during the EoR. Starting from a suite of six $50h^{-1}$~cMpc \texttt{dustyGadget} simulations, we provided indications on the most promising galaxy candidates with $M_\star > 10^{7.5} ~ \si{M_\odot}$ to look for PISNe at $6 \lesssim z \lesssim 8$. We also provided predictions on the expected number of PISNe in selected JWST surveys and in a future $\sim 1$~deg$^2$ survey with the Nancy Grace Roman Space Telescope. Finally, we discussed the observability of PISNe within their hosting galaxies in these surveys, in terms of both their integrated and resolved stellar emission. We considered many possible sources of uncertainties in our results, including the effect of mass loss on Pop\ III lifetimes, the impact of choosing different IMFs on the expected number of PISNe per unit Pop\ III mass, and the possibility of a reduced Pop\ III star formation efficiency with respect to the nominal efficiency of the simulation.

We find that, although the probability of observing PISNe is indeed small, it is non-negliglible. In our reference model for Pop\ III (Salpeter-like IMF in the range [100, 500]~\si{M_\odot}, Pop\ III star formation efficiency of $\eta_\mathrm{III} = 0.1$ and no mass loss), the comoving number density of PISNe and the average number of PISNe per halo can be up to $\sim 10^{-1} ~ \si{cMpc^{-3}}$ and $\sim 5 \times 10^{-6}$ (i.e. about 1 PISN every two hundred thousand haloes, on average) respectively. A higher number density of PISNe is generally found in haloes with decreasing stellar mass, given the higher absolute number density of low-mass haloes at all the considered redshifts. However, the relative fraction of low-mass haloes hosting PISNe is smaller with respect to higher-mass haloes, and hence the probability of finding PISNe in a given halo increases with $M_\star$, and it is highest in haloes with $10^{9.5} ~ \si{M_\odot} \lesssim M_\star \lesssim 10^{10} ~ \si{M_\odot}$ at $z = 6.7$. Therefore, different observing strategies may be considered: either targeted follow-up observations of candidate high-mass galaxies - with a higher PISNe-hosting fraction -, or blind wide-field surveys - also including the numerous low-mass galaxies that can also potentially host PISNe.

PISNe from progenitors with masses higher than 200~\si{M_\odot} would be observable for a variable time range (from $\sim$~2~months up to $\sim$~9~yr) through all the considered JWST/NIRCam filters, particularly the F356W and F200W filters that have been indicated by \citealt{Hartwig_2018} as the best two-filter diagnostic to identify possible PISNe candidates, and the F444W and F277W filters available for the largest considered JWST survey (COSMOS-Web). The 150~\si{M_\odot}-progenitor PISN would also be visible for almost one/two years in the F200W/F356W filters of the deepest considered JWST survey (NGDEEP). However, detecting PISNe with JWST may still be challenging due to their rarity: in our reference model, we expect on average less than 1 PISN in all the examined JWST surveys. 

A higher potential might be obtained through the WFI instrument on board of Roman, thanks to its large field-of-view. By considering an example $\sim 1$~deg$^2$ survey, we see that $\simeq 1.5 \, \eta_\mathrm{III}$ PISNe are expected in our reference model. An even higher survey volume would further increase the number of expected PISNe. For example, the $\sim 10$~deg$^2$ survey suggested by \citealt{Moriya_2022_RST} with limiting magnitudes of 27.0~mag and 26.5~mag in the F158 and F213 filters (low enough to see at least the 250~\si{M_\odot}-progenitor PISN for about a year and a half/three years in F158/F213) would further increase this number by a factor ten. More favourable scenarios are also obtained when considering different Pop\ III IMFs and/or higher Pop\ III formation efficiencies, and when including the contribution of coarsely-resolved environments that have not been specifically targeted in this study.

While the integrated flux of the underlying galaxies might exceed the flux of PISNe in the considered filters, Pop\ III stars - and hence PISNe - are usually found at the periphery of their hosting haloes. This should mitigate the contamination of the signal arising from stars, especially for the most massive galaxies in which the stellar emission is more likely to outshine PISNe. The bulk of most PISN-hosting galaxies would also be resolved in JWST/NIRCam and Roman/WFI at these redshifts.

We remark once again that if we do find a galaxy hosting a PISN at high redshifts, this incredible accomplishment may also pave the way for an even more incredible discovery. Indeed, by observing the target galaxy again after some years, when the SN has faded considerably, we might even be able to discern an active underlying Pop\ III stellar population. This would possibly lead to the first clear detection of Pop\ III stars in the history of astronomy.

\section*{Acknowledgments}
LG and RS acknowledge support from the Amaldi Research Center funded by the Italian Ministry for Education, Universities and Research (MIUR) program "Dipartimento di Eccellenza" (CUP:B81I18001170001). We have benefited from the publicly available programming language \texttt{Python}, including the \texttt{numpy}, \texttt{matplotlib} and \texttt{scipy} packages.

\section*{Data Availability}
The data underlying this article will be shared on reasonable request to the corresponding author.

\bibliographystyle{mn2e}
\bibliography{main}

\begin{thebibliography}{}
\makeatletter
\relax
\def\mn@urlcharsother{\let\do\@makeother \do\$\do\&\do\#\do\^\do\_\do\%\do\~}
\def\mn@doi{\begingroup\mn@urlcharsother \@ifnextchar [ {\mn@doi@} {\mn@doi@[]}}
\def\mn@doi@[#1]#2{\def\@tempa{#1}\ifx\@tempa\@empty \href {http://dx.doi.org/#2} {doi:#2}\else \href {http://dx.doi.org/#2} {#1}\fi \endgroup}
\def\mn@eprint#1#2{\mn@eprint@#1:#2::\@nil}
\def\mn@eprint@arXiv#1{\href {http://arxiv.org/abs/#1} {{\tt arXiv:#1}}}
\def\mn@eprint@dblp#1{\href {http://dblp.uni-trier.de/rec/bibtex/#1.xml} {dblp:#1}}
\def\mn@eprint@#1:#2:#3:#4\@nil{\def\@tempa {#1}\def\@tempb {#2}\def\@tempc {#3}\ifx \@tempc \@empty \let \@tempc \@tempb \let \@tempb \@tempa \fi \ifx \@tempb \@empty \def\@tempb {arXiv}\fi \@ifundefined {mn@eprint@\@tempb}{\@tempb:\@tempc}{\expandafter \expandafter \csname mn@eprint@\@tempb\endcsname \expandafter{\@tempc}}}

\bibitem[\protect\citeauthoryear{{Abel}, {Bryan}  \& {Norman}}{{Abel} et~al.}{2002}]{Abel_2002}
{Abel} T.,  {Bryan} G.~L.,   {Norman} M.~L.,  2002, \mn@doi [Science] {10.1126/science.295.5552.93}, \href {https://ui.adsabs.harvard.edu/abs/2002Sci...295...93A} {295, 93}

\bibitem[\protect\citeauthoryear{{Aguado} et~al.,}{{Aguado} et~al.}{2023a}]{Aguado_2023_PISN-explorer}
{Aguado} D.~S.,  et~al., 2023a, \mn@doi [\mnras] {10.1093/mnras/stad164}, \href {https://ui.adsabs.harvard.edu/abs/2023MNRAS.520..866A} {520, 866}

\bibitem[\protect\citeauthoryear{{Aguado} et~al.,}{{Aguado} et~al.}{2023b}]{Aguado_2023_mostMetalPoorStar}
{Aguado} D.~S.,  et~al., 2023b, \mn@doi [\aap] {10.1051/0004-6361/202245392}, \href {https://ui.adsabs.harvard.edu/abs/2023A&A...669L...4A} {669, L4}

\bibitem[\protect\citeauthoryear{{Anders} \& {Grevesse}}{{Anders} \& {Grevesse}}{1989}]{Anders_Grevesse_1989}
{Anders} E.,  {Grevesse} N.,  1989, \mn@doi [\gca] {10.1016/0016-7037(89)90286-X}, \href {https://ui.adsabs.harvard.edu/abs/1989GeCoA..53..197A} {53, 197}

\bibitem[\protect\citeauthoryear{{Aoki}, {Tominaga}, {Beers}, {Honda}  \& {Lee}}{{Aoki} et~al.}{2014}]{Aoki_2014}
{Aoki} W.,  {Tominaga} N.,  {Beers} T.~C.,  {Honda} S.,   {Lee} Y.~S.,  2014, \mn@doi [Science] {10.1126/science.1252633}, \href {https://ui.adsabs.harvard.edu/abs/2014Sci...345..912A} {345, 912}

\bibitem[\protect\citeauthoryear{{Bagley} et~al.,}{{Bagley} et~al.}{2023}]{Bagley_2023}
{Bagley} M.~B.,  et~al., 2023, \mn@doi [arXiv e-prints] {10.48550/arXiv.2302.05466}, \href {https://ui.adsabs.harvard.edu/abs/2023arXiv230205466B} {p. arXiv:2302.05466}

\bibitem[\protect\citeauthoryear{{Banik}, {Tan}  \& {Monaco}}{{Banik} et~al.}{2019}]{Banik_2019}
{Banik} N.,  {Tan} J.~C.,   {Monaco} P.,  2019, \mn@doi [\mnras] {10.1093/mnras/sty3298}, \href {https://ui.adsabs.harvard.edu/abs/2019MNRAS.483.3592B} {483, 3592}

\bibitem[\protect\citeauthoryear{{Baraffe}, {Heger}  \& {Woosley}}{{Baraffe} et~al.}{2001}]{Baraffe_2001}
{Baraffe} I.,  {Heger} A.,   {Woosley} S.~E.,  2001, \mn@doi [\apj] {10.1086/319808}, \href {https://ui.adsabs.harvard.edu/abs/2001ApJ...550..890B} {550, 890}

\bibitem[\protect\citeauthoryear{{Barkat}, {Rakavy}  \& {Sack}}{{Barkat} et~al.}{1967}]{Barkat_1967}
{Barkat} Z.,  {Rakavy} G.,   {Sack} N.,  1967, \mn@doi [\prl] {10.1103/PhysRevLett.18.379}, \href {https://ui.adsabs.harvard.edu/abs/1967PhRvL..18..379B} {18, 379}

\bibitem[\protect\citeauthoryear{{Bennett} \& {Sijacki}}{{Bennett} \& {Sijacki}}{2020}]{Bennet_Sijacki_2020}
{Bennett} J.~S.,  {Sijacki} D.,  2020, \mn@doi [\mnras] {10.1093/mnras/staa2835}, \href {https://ui.adsabs.harvard.edu/abs/2020MNRAS.499..597B} {499, 597}

\bibitem[\protect\citeauthoryear{{Bezanson} et~al.,}{{Bezanson} et~al.}{2022}]{Bezanson_2022}
{Bezanson} R.,  et~al., 2022, \mn@doi [arXiv e-prints] {10.48550/arXiv.2212.04026}, \href {https://ui.adsabs.harvard.edu/abs/2022arXiv221204026B} {p. arXiv:2212.04026}

\bibitem[\protect\citeauthoryear{{Bond}, {Arnett}  \& {Carr}}{{Bond} et~al.}{1984}]{Bond_1984}
{Bond} J.~R.,  {Arnett} W.~D.,   {Carr} B.~J.,  1984, \mn@doi [\apj] {10.1086/162057}, \href {https://ui.adsabs.harvard.edu/abs/1984ApJ...280..825B} {280, 825}

\bibitem[\protect\citeauthoryear{{Bromm}}{{Bromm}}{2013}]{Bromm_2013}
{Bromm} V.,  2013, \mn@doi [Reports on Progress in Physics] {10.1088/0034-4885/76/11/112901}, \href {https://ui.adsabs.harvard.edu/abs/2013RPPh...76k2901B} {76, 112901}

\bibitem[\protect\citeauthoryear{{Bromm}, {Ferrara}, {Coppi}  \& {Larson}}{{Bromm} et~al.}{2001a}]{Bromm_2001}
{Bromm} V.,  {Ferrara} A.,  {Coppi} P.~S.,   {Larson} R.~B.,  2001a, \mn@doi [\mnras] {10.1046/j.1365-8711.2001.04915.x}, \href {https://ui.adsabs.harvard.edu/abs/2001MNRAS.328..969B} {328, 969}

\bibitem[\protect\citeauthoryear{{Bromm}, {Kudritzki}  \& {Loeb}}{{Bromm} et~al.}{2001b}]{Bromm_2001_spectra}
{Bromm} V.,  {Kudritzki} R.~P.,   {Loeb} A.,  2001b, \mn@doi [\apj] {10.1086/320549}, \href {https://ui.adsabs.harvard.edu/abs/2001ApJ...552..464B} {552, 464}

\bibitem[\protect\citeauthoryear{{Bromm}, {Coppi}  \& {Larson}}{{Bromm} et~al.}{2002}]{Bromm_2002}
{Bromm} V.,  {Coppi} P.~S.,   {Larson} R.~B.,  2002, \mn@doi [\apj] {10.1086/323947}, \href {https://ui.adsabs.harvard.edu/abs/2002ApJ...564...23B} {564, 23}

\bibitem[\protect\citeauthoryear{{Casey} et~al.,}{{Casey} et~al.}{2023}]{Casey_2023}
{Casey} C.~M.,  et~al., 2023, \mn@doi [\apj] {10.3847/1538-4357/acc2bc}, \href {https://ui.adsabs.harvard.edu/abs/2023ApJ...954...31C} {954, 31}

\bibitem[\protect\citeauthoryear{{Castellano} et~al.,}{{Castellano} et~al.}{2022}]{Castellano_2022}
{Castellano} M.,  et~al., 2022, \mn@doi [\apjl] {10.3847/2041-8213/ac94d0}, \href {https://ui.adsabs.harvard.edu/abs/2022ApJ...938L..15C} {938, L15}

\bibitem[\protect\citeauthoryear{{Chen}, {Heger}, {Woosley}, {Almgren}  \& {Whalen}}{{Chen} et~al.}{2014}]{Chen_2014}
{Chen} K.-J.,  {Heger} A.,  {Woosley} S.,  {Almgren} A.,   {Whalen} D.~J.,  2014, \mn@doi [\apj] {10.1088/0004-637X/792/1/44}, \href {https://ui.adsabs.harvard.edu/abs/2014ApJ...792...44C} {792, 44}

\bibitem[\protect\citeauthoryear{{Chiaki} \& {Yoshida}}{{Chiaki} \& {Yoshida}}{2022}]{Chiaki_Yoshida_2022}
{Chiaki} G.,  {Yoshida} N.,  2022, \mn@doi [\mnras] {10.1093/mnras/stab2799}, \href {https://ui.adsabs.harvard.edu/abs/2022MNRAS.510.5199C} {510, 5199}

\bibitem[\protect\citeauthoryear{{Chiaki}, {Yoshida}  \& {Hirano}}{{Chiaki} et~al.}{2016}]{Chiaki_2016}
{Chiaki} G.,  {Yoshida} N.,   {Hirano} S.,  2016, \mn@doi [\mnras] {10.1093/mnras/stw2120}, \href {https://ui.adsabs.harvard.edu/abs/2016MNRAS.463.2781C} {463, 2781}

\bibitem[\protect\citeauthoryear{{Chon}, {Omukai}  \& {Schneider}}{{Chon} et~al.}{2021}]{Chon_2021}
{Chon} S.,  {Omukai} K.,   {Schneider} R.,  2021, \mn@doi [\mnras] {10.1093/mnras/stab2497}, \href {https://ui.adsabs.harvard.edu/abs/2021MNRAS.508.4175C} {508, 4175}

\bibitem[\protect\citeauthoryear{{Chon}, {Ono}, {Omukai}  \& {Schneider}}{{Chon} et~al.}{2022}]{Chon_2022}
{Chon} S.,  {Ono} H.,  {Omukai} K.,   {Schneider} R.,  2022, \mn@doi [\mnras] {10.1093/mnras/stac1549}, \href {https://ui.adsabs.harvard.edu/abs/2022MNRAS.514.4639C} {514, 4639}

\bibitem[\protect\citeauthoryear{{Cleri} et~al.,}{{Cleri} et~al.}{2023}]{Cleri_2023}
{Cleri} N.~J.,  et~al., 2023, \mn@doi [\apj] {10.3847/1538-4357/acde55}, \href {https://ui.adsabs.harvard.edu/abs/2023ApJ...953...10C} {953, 10}

\bibitem[\protect\citeauthoryear{{Cooke} et~al.,}{{Cooke} et~al.}{2012}]{Cooke_2012}
{Cooke} J.,  et~al., 2012, \mn@doi [\nat] {10.1038/nature11521}, \href {https://ui.adsabs.harvard.edu/abs/2012Natur.491..228C} {491, 228}

\bibitem[\protect\citeauthoryear{{Dessart}, {Hillier}, {Waldman}, {Livne}  \& {Blondin}}{{Dessart} et~al.}{2012}]{Dessart_2012}
{Dessart} L.,  {Hillier} D.~J.,  {Waldman} R.,  {Livne} E.,   {Blondin} S.,  2012, \mn@doi [\mnras] {10.1111/j.1745-3933.2012.01329.x}, \href {https://ui.adsabs.harvard.edu/abs/2012MNRAS.426L..76D} {426, L76}

\bibitem[\protect\citeauthoryear{{Di Cesare}, {Graziani}, {Schneider}, {Ginolfi}, {Venditti}, {Santini}  \& {Hunt}}{{Di Cesare} et~al.}{2023}]{DiCesare_2022}
{Di Cesare} C.,  {Graziani} L.,  {Schneider} R.,  {Ginolfi} M.,  {Venditti} A.,  {Santini} P.,   {Hunt} L.~K.,  2023, \mn@doi [\mnras] {10.1093/mnras/stac370210.48550/arXiv.2209.05496}, \href {https://ui.adsabs.harvard.edu/abs/2023MNRAS.519.4632D} {519, 4632}

\bibitem[\protect\citeauthoryear{{Dunlop} et~al.,}{{Dunlop} et~al.}{2021}]{Dunlop_2021}
{Dunlop} J.~S.,  et~al., 2021, {PRIMER: Public Release IMaging for Extragalactic Research}, JWST Proposal. Cycle 1, ID. \#1837

\bibitem[\protect\citeauthoryear{{Duric}}{{Duric}}{2003}]{Duric_2003}
{Duric} N.,  2003, {Advanced Astrophysics}

\bibitem[\protect\citeauthoryear{{Euclid Collaboration} et~al.,}{{Euclid Collaboration} et~al.}{2022}]{EuclidCollaboration_2022}
{Euclid Collaboration} et~al., 2022, \mn@doi [\aap] {10.1051/0004-6361/202142897}, \href {https://ui.adsabs.harvard.edu/abs/2022A&A...662A..92E} {662, A92}

\bibitem[\protect\citeauthoryear{{Finkelstein} et~al.,}{{Finkelstein} et~al.}{2017}]{Finkelstein_2017}
{Finkelstein} S.~L.,  et~al., 2017, {The Cosmic Evolution Early Release Science (CEERS) Survey}, JWST Proposal ID 1345. Cycle 0 Early Release Science

\bibitem[\protect\citeauthoryear{{Finkelstein} et~al.,}{{Finkelstein} et~al.}{2021}]{Finkelstein_2021}
{Finkelstein} S.~L.,  et~al., 2021, {The Next Generation Deep Extragalactic Exploratory Public (NGDEEP) Survey: Feedback in Low-Mass Galaxies from Cosmic Dawn to Dusk}, JWST Proposal. Cycle 1, ID. \#2079

\bibitem[\protect\citeauthoryear{{Finkelstein} et~al.,}{{Finkelstein} et~al.}{2022}]{Finkelstein_2022}
{Finkelstein} S.~L.,  et~al., 2022, \mn@doi [\apjl] {10.3847/2041-8213/ac966e}, \href {https://ui.adsabs.harvard.edu/abs/2022ApJ...940L..55F} {940, L55}

\bibitem[\protect\citeauthoryear{{Finkelstein} et~al.,}{{Finkelstein} et~al.}{2023}]{Finkelstein_2023}
{Finkelstein} S.~L.,  et~al., 2023, \mn@doi [\apjl] {10.3847/2041-8213/acade4}, \href {https://ui.adsabs.harvard.edu/abs/2023ApJ...946L..13F} {946, L13}

\bibitem[\protect\citeauthoryear{{Fraley}}{{Fraley}}{1968}]{Fraley_1968}
{Fraley} G.~S.,  1968, \mn@doi [\apss] {10.1007/BF00651498}, \href {https://ui.adsabs.harvard.edu/abs/1968Ap&SS...2...96F} {2, 96}

\bibitem[\protect\citeauthoryear{{Fraser}, {Casey}, {Gilmore}, {Heger}  \& {Chan}}{{Fraser} et~al.}{2017}]{Fraser_2017}
{Fraser} M.,  {Casey} A.~R.,  {Gilmore} G.,  {Heger} A.,   {Chan} C.,  2017, \mn@doi [\mnras] {10.1093/mnras/stx480}, \href {https://ui.adsabs.harvard.edu/abs/2017MNRAS.468..418F} {468, 418}

\bibitem[\protect\citeauthoryear{{Freese}, {Bodenheimer}, {Spolyar}  \& {Gondolo}}{{Freese} et~al.}{2008}]{Freese_2008}
{Freese} K.,  {Bodenheimer} P.,  {Spolyar} D.,   {Gondolo} P.,  2008, \mn@doi [\apjl] {10.1086/592685}, \href {https://ui.adsabs.harvard.edu/abs/2008ApJ...685L.101F} {685, L101}

\bibitem[\protect\citeauthoryear{{Freese}, {Ilie}, {Spolyar}, {Valluri}  \& {Bodenheimer}}{{Freese} et~al.}{2010}]{Freese_2010}
{Freese} K.,  {Ilie} C.,  {Spolyar} D.,  {Valluri} M.,   {Bodenheimer} P.,  2010, \mn@doi [\apj] {10.1088/0004-637X/716/2/1397}, \href {https://ui.adsabs.harvard.edu/abs/2010ApJ...716.1397F} {716, 1397}

\bibitem[\protect\citeauthoryear{{Fryer}, {Woosley}  \& {Heger}}{{Fryer} et~al.}{2001}]{Fryer_2001}
{Fryer} C.~L.,  {Woosley} S.~E.,   {Heger} A.,  2001, \mn@doi [\apj] {10.1086/319719}, \href {https://ui.adsabs.harvard.edu/abs/2001ApJ...550..372F} {550, 372}

\bibitem[\protect\citeauthoryear{{Furtak} et~al.,}{{Furtak} et~al.}{2023}]{Furtak_2023}
{Furtak} L.~J.,  et~al., 2023, \mn@doi [\mnras] {10.1093/mnras/stad1627}, \href {https://ui.adsabs.harvard.edu/abs/2023MNRAS.523.4568F} {523, 4568}

\bibitem[\protect\citeauthoryear{{Gal-Yam} et~al.,}{{Gal-Yam} et~al.}{2009}]{Gal-Yam_2008}
{Gal-Yam} A.,  et~al., 2009, \mn@doi [\nat] {10.1038/nature08579}, \href {https://ui.adsabs.harvard.edu/abs/2009Natur.462..624G} {462, 624}

\bibitem[\protect\citeauthoryear{{Gilmer}, {Kozyreva}, {Hirschi}, {Fr{\"o}hlich}  \& {Yusof}}{{Gilmer} et~al.}{2017}]{Gilmer_2017}
{Gilmer} M.~S.,  {Kozyreva} A.,  {Hirschi} R.,  {Fr{\"o}hlich} C.,   {Yusof} N.,  2017, \mn@doi [\apj] {10.3847/1538-4357/aa8461}, \href {https://ui.adsabs.harvard.edu/abs/2017ApJ...846..100G} {846, 100}

\bibitem[\protect\citeauthoryear{{Ginolfi} et~al.,}{{Ginolfi} et~al.}{2020}]{Ginolfi_2020}
{Ginolfi} M.,  et~al., 2020, \mn@doi [\aap] {10.1051/0004-6361/201936872}, \href {https://ui.adsabs.harvard.edu/abs/2020A&A...633A..90G} {633, A90}

\bibitem[\protect\citeauthoryear{{Graziani}, {Schneider}, {Ginolfi}, {Hunt}, {Maio}, {Glatzle}  \& {Ciardi}}{{Graziani} et~al.}{2020}]{Graziani_2020}
{Graziani} L.,  {Schneider} R.,  {Ginolfi} M.,  {Hunt} L.~K.,  {Maio} U.,  {Glatzle} M.,   {Ciardi} B.,  2020, \mn@doi [\mnras] {10.1093/mnras/staa79610.48550/arXiv.1909.07388}, \href {https://ui.adsabs.harvard.edu/abs/2020MNRAS.494.1071G} {494, 1071}

\bibitem[\protect\citeauthoryear{{Greif}, {Glover}, {Bromm}  \& {Klessen}}{{Greif} et~al.}{2010}]{Greif_2010}
{Greif} T.~H.,  {Glover} S. C.~O.,  {Bromm} V.,   {Klessen} R.~S.,  2010, \mn@doi [\apj] {10.1088/0004-637X/716/1/510}, \href {https://ui.adsabs.harvard.edu/abs/2010ApJ...716..510G} {716, 510}

\bibitem[\protect\citeauthoryear{{Haardt} \& {Madau}}{{Haardt} \& {Madau}}{1996}]{Haardt_Madau_1996}
{Haardt} F.,  {Madau} P.,  1996, \mn@doi [\apj] {10.1086/177035}, \href {https://ui.adsabs.harvard.edu/abs/1996ApJ...461...20H} {461, 20}

\bibitem[\protect\citeauthoryear{{Haemmerl{\'e}}, {Mayer}, {Klessen}, {Hosokawa}, {Madau}  \& {Bromm}}{{Haemmerl{\'e}} et~al.}{2020}]{Hammerle_2020}
{Haemmerl{\'e}} L.,  {Mayer} L.,  {Klessen} R.~S.,  {Hosokawa} T.,  {Madau} P.,   {Bromm} V.,  2020, \mn@doi [\ssr] {10.1007/s11214-020-00673-y}, \href {https://ui.adsabs.harvard.edu/abs/2020SSRv..216...48H} {216, 48}

\bibitem[\protect\citeauthoryear{{Harikane}, {Nakajima}, {Ouchi}, {Umeda}, {Isobe}, {Ono}, {Xu}  \& {Zhang}}{{Harikane} et~al.}{2023}]{Harikane_2023}
{Harikane} Y.,  {Nakajima} K.,  {Ouchi} M.,  {Umeda} H.,  {Isobe} Y.,  {Ono} Y.,  {Xu} Y.,   {Zhang} Y.,  2023, \mn@doi [arXiv e-prints] {10.48550/arXiv.2304.06658}, \href {https://ui.adsabs.harvard.edu/abs/2023arXiv230406658H} {p. arXiv:2304.06658}

\bibitem[\protect\citeauthoryear{{Hartwig}, {Bromm}, {Klessen}  \& {Glover}}{{Hartwig} et~al.}{2015}]{Hartwig_2015}
{Hartwig} T.,  {Bromm} V.,  {Klessen} R.~S.,   {Glover} S. C.~O.,  2015, \mn@doi [\mnras] {10.1093/mnras/stu2740}, \href {https://ui.adsabs.harvard.edu/abs/2015MNRAS.447.3892H} {447, 3892}

\bibitem[\protect\citeauthoryear{{Hartwig} et~al.,}{{Hartwig} et~al.}{2018}]{Hartwig_2018}
{Hartwig} T.,  et~al., 2018, \mn@doi [\mnras] {10.1093/mnras/sty1176}, \href {https://ui.adsabs.harvard.edu/abs/2018MNRAS.478.1795H} {478, 1795}

\bibitem[\protect\citeauthoryear{{Heger} \& {Woosley}}{{Heger} \& {Woosley}}{2002}]{Heger_Woosley_2002}
{Heger} A.,  {Woosley} S.~E.,  2002, \mn@doi [\apj] {10.1086/338487}, \href {https://ui.adsabs.harvard.edu/abs/2002ApJ...567..532H} {567, 532}

\bibitem[\protect\citeauthoryear{{Hirano}, {Umeda}  \& {Yoshida}}{{Hirano} et~al.}{2011}]{Hirano_2011}
{Hirano} S.,  {Umeda} H.,   {Yoshida} N.,  2011, \mn@doi [\apj] {10.1088/0004-637X/736/1/58}, \href {https://ui.adsabs.harvard.edu/abs/2011ApJ...736...58H} {736, 58}

\bibitem[\protect\citeauthoryear{{Hirano}, {Hosokawa}, {Yoshida}, {Umeda}, {Omukai}, {Chiaki}  \& {Yorke}}{{Hirano} et~al.}{2014}]{Hirano_2014}
{Hirano} S.,  {Hosokawa} T.,  {Yoshida} N.,  {Umeda} H.,  {Omukai} K.,  {Chiaki} G.,   {Yorke} H.~W.,  2014, \mn@doi [\apj] {10.1088/0004-637X/781/2/60}, \href {https://ui.adsabs.harvard.edu/abs/2014ApJ...781...60H} {781, 60}

\bibitem[\protect\citeauthoryear{{Hirano}, {Hosokawa}, {Yoshida}, {Omukai}  \& {Yorke}}{{Hirano} et~al.}{2015a}]{Hirano_2015_UVrad}
{Hirano} S.,  {Hosokawa} T.,  {Yoshida} N.,  {Omukai} K.,   {Yorke} H.~W.,  2015a, \mn@doi [\mnras] {10.1093/mnras/stv044}, \href {https://ui.adsabs.harvard.edu/abs/2015MNRAS.448..568H} {448, 568}

\bibitem[\protect\citeauthoryear{{Hirano}, {Zhu}, {Yoshida}, {Spergel}  \& {Yorke}}{{Hirano} et~al.}{2015b}]{Hirano_2015_PPS}
{Hirano} S.,  {Zhu} N.,  {Yoshida} N.,  {Spergel} D.,   {Yorke} H.~W.,  2015b, \mn@doi [\apj] {10.1088/0004-637X/814/1/18}, \href {https://ui.adsabs.harvard.edu/abs/2015ApJ...814...18H} {814, 18}

\bibitem[\protect\citeauthoryear{{Hosokawa}, {Omukai}, {Yoshida}  \& {Yorke}}{{Hosokawa} et~al.}{2011}]{Hosokawa_2011}
{Hosokawa} T.,  {Omukai} K.,  {Yoshida} N.,   {Yorke} H.~W.,  2011, \mn@doi [Science] {10.1126/science.1207433}, \href {https://ui.adsabs.harvard.edu/abs/2011Sci...334.1250H} {334, 1250}

\bibitem[\protect\citeauthoryear{{Hosokawa}, {Hirano}, {Kuiper}, {Yorke}, {Omukai}  \& {Yoshida}}{{Hosokawa} et~al.}{2016}]{Hosokawa_2016}
{Hosokawa} T.,  {Hirano} S.,  {Kuiper} R.,  {Yorke} H.~W.,  {Omukai} K.,   {Yoshida} N.,  2016, \mn@doi [\apj] {10.3847/0004-637X/824/2/119}, \href {https://ui.adsabs.harvard.edu/abs/2016ApJ...824..119H} {824, 119}

\bibitem[\protect\citeauthoryear{{Hummel}, {Pawlik}, {Milosavljevi{\'c}}  \& {Bromm}}{{Hummel} et~al.}{2012}]{Hummel_2012}
{Hummel} J.~A.,  {Pawlik} A.~H.,  {Milosavljevi{\'c}} M.,   {Bromm} V.,  2012, \mn@doi [\apj] {10.1088/0004-637X/755/1/72}, \href {https://ui.adsabs.harvard.edu/abs/2012ApJ...755...72H} {755, 72}

\bibitem[\protect\citeauthoryear{{Inoue}}{{Inoue}}{2011}]{Inoue_2011}
{Inoue} A.~K.,  2011, \mn@doi [\mnras] {10.1111/j.1365-2966.2011.18906.x}, \href {https://ui.adsabs.harvard.edu/abs/2011MNRAS.415.2920I} {415, 2920}

\bibitem[\protect\citeauthoryear{{Ishigaki}, {Tominaga}, {Kobayashi}  \& {Nomoto}}{{Ishigaki} et~al.}{2014}]{Ishigaki_2014}
{Ishigaki} M.~N.,  {Tominaga} N.,  {Kobayashi} C.,   {Nomoto} K.,  2014, \mn@doi [\apjl] {10.1088/2041-8205/792/2/L32}, \href {https://ui.adsabs.harvard.edu/abs/2014ApJ...792L..32I} {792, L32}

\bibitem[\protect\citeauthoryear{{Iwamoto}, {Umeda}, {Tominaga}, {Nomoto}  \& {Maeda}}{{Iwamoto} et~al.}{2005}]{2Iwamoto_2005}
{Iwamoto} N.,  {Umeda} H.,  {Tominaga} N.,  {Nomoto} K.,   {Maeda} K.,  2005, \mn@doi [Science] {10.1126/science.1112997}, \href {https://ui.adsabs.harvard.edu/abs/2005Sci...309..451I} {309, 451}

\bibitem[\protect\citeauthoryear{{Jaacks}, {Thompson}, {Finkelstein}  \& {Bromm}}{{Jaacks} et~al.}{2018}]{Jaacks_2018_metalEnrichment}
{Jaacks} J.,  {Thompson} R.,  {Finkelstein} S.~L.,   {Bromm} V.,  2018, \mn@doi [\mnras] {10.1093/mnras/sty062}, \href {https://ui.adsabs.harvard.edu/abs/2018MNRAS.475.4396J} {475, 4396}

\bibitem[\protect\citeauthoryear{{Jaacks}, {Finkelstein}  \& {Bromm}}{{Jaacks} et~al.}{2019}]{Jaacks_2019}
{Jaacks} J.,  {Finkelstein} S.~L.,   {Bromm} V.,  2019, \mn@doi [\mnras] {10.1093/mnras/stz1529}, \href {https://ui.adsabs.harvard.edu/abs/2019MNRAS.488.2202J} {488, 2202}

\bibitem[\protect\citeauthoryear{{Jeena}, {Banerjee}  \& {Heger}}{{Jeena} et~al.}{2023}]{Jeena_2023}
{Jeena} S.~K.,  {Banerjee} P.,   {Heger} A.,  2023, \mn@doi [arXiv e-prints] {10.48550/arXiv.2310.00591}, \href {https://ui.adsabs.harvard.edu/abs/2023arXiv231000591J} {p. arXiv:2310.00591}

\bibitem[\protect\citeauthoryear{{Jeon}, {Bromm}, {Pawlik}  \& {Milosavljevi{\'c}}}{{Jeon} et~al.}{2015}]{Jeon_2015}
{Jeon} M.,  {Bromm} V.,  {Pawlik} A.~H.,   {Milosavljevi{\'c}} M.,  2015, \mn@doi [\mnras] {10.1093/mnras/stv1353}, \href {https://ui.adsabs.harvard.edu/abs/2015MNRAS.452.1152J} {452, 1152}

\bibitem[\protect\citeauthoryear{{Jeon}, {Besla}  \& {Bromm}}{{Jeon} et~al.}{2017}]{Jeon_2017}
{Jeon} M.,  {Besla} G.,   {Bromm} V.,  2017, \mn@doi [\apj] {10.3847/1538-4357/aa8c80}, \href {https://ui.adsabs.harvard.edu/abs/2017ApJ...848...85J} {848, 85}

\bibitem[\protect\citeauthoryear{{Jerkstrand}, {Smartt}  \& {Heger}}{{Jerkstrand} et~al.}{2016}]{Jerkstrand_2016}
{Jerkstrand} A.,  {Smartt} S.~J.,   {Heger} A.,  2016, \mn@doi [\mnras] {10.1093/mnras/stv2369}, \href {https://ui.adsabs.harvard.edu/abs/2016MNRAS.455.3207J} {455, 3207}

\bibitem[\protect\citeauthoryear{{Ji}, {Frebel}  \& {Bromm}}{{Ji} et~al.}{2015}]{Ji_2015}
{Ji} A.~P.,  {Frebel} A.,   {Bromm} V.,  2015, \mn@doi [\mnras] {10.1093/mnras/stv2052}, \href {https://ui.adsabs.harvard.edu/abs/2015MNRAS.454..659J} {454, 659}

\bibitem[\protect\citeauthoryear{{Johnson}, {Dalla Vecchia}  \& {Khochfar}}{{Johnson} et~al.}{2013}]{Johnson_2013}
{Johnson} J.~L.,  {Dalla Vecchia} C.,   {Khochfar} S.,  2013, \mn@doi [\mnras] {10.1093/mnras/sts011}, \href {https://ui.adsabs.harvard.edu/abs/2013MNRAS.428.1857J} {428, 1857}

\bibitem[\protect\citeauthoryear{{Jones} et~al.,}{{Jones} et~al.}{2013}]{Jones_2013}
{Jones} D.~O.,  et~al., 2013, \mn@doi [\apj] {10.1088/0004-637X/768/2/166}, \href {https://ui.adsabs.harvard.edu/abs/2013ApJ...768..166J} {768, 166}

\bibitem[\protect\citeauthoryear{{Karlsson}, {Johnson}  \& {Bromm}}{{Karlsson} et~al.}{2008}]{Karlsson_2008}
{Karlsson} T.,  {Johnson} J.~L.,   {Bromm} V.,  2008, \mn@doi [\apj] {10.1086/533520}, \href {https://ui.adsabs.harvard.edu/abs/2008ApJ...679....6K} {679, 6}

\bibitem[\protect\citeauthoryear{{Karlsson}, {Bromm}  \& {Bland-Hawthorn}}{{Karlsson} et~al.}{2013}]{Karlsson_2013}
{Karlsson} T.,  {Bromm} V.,   {Bland-Hawthorn} J.,  2013, \mn@doi [Reviews of Modern Physics] {10.1103/RevModPhys.85.809}, \href {https://ui.adsabs.harvard.edu/abs/2013RvMP...85..809K} {85, 809}

\bibitem[\protect\citeauthoryear{{Kasen}, {Woosley}  \& {Heger}}{{Kasen} et~al.}{2011}]{Kasen_2011}
{Kasen} D.,  {Woosley} S.~E.,   {Heger} A.,  2011, \mn@doi [\apj] {10.1088/0004-637X/734/2/102}, \href {https://ui.adsabs.harvard.edu/abs/2011ApJ...734..102K} {734, 102}

\bibitem[\protect\citeauthoryear{{Katz}, {Kimm}, {Ellis}, {Devriendt}  \& {Slyz}}{{Katz} et~al.}{2023}]{Katz_2023}
{Katz} H.,  {Kimm} T.,  {Ellis} R.~S.,  {Devriendt} J.,   {Slyz} A.,  2023, \mn@doi [\mnras] {10.1093/mnras/stad1903}, \href {https://ui.adsabs.harvard.edu/abs/2023MNRAS.524..351K} {524, 351}

\bibitem[\protect\citeauthoryear{{Keller} et~al.,}{{Keller} et~al.}{2014}]{Keller_2014}
{Keller} S.~C.,  et~al., 2014, \mn@doi [\nat] {10.1038/nature12990}, \href {https://ui.adsabs.harvard.edu/abs/2014Natur.506..463K} {506, 463}

\bibitem[\protect\citeauthoryear{{Kippenhahn}, {Weigert}  \& {Weiss}}{{Kippenhahn} et~al.}{2013}]{Kippenhahn_2013}
{Kippenhahn} R.,  {Weigert} A.,   {Weiss} A.,  2013, {Stellar Structure and Evolution}, \mn@doi{10.1007/978-3-642-30304-3.
}

\bibitem[\protect\citeauthoryear{{Klessen} \& {Glover}}{{Klessen} \& {Glover}}{2023}]{Klessen_Glover_2023}
{Klessen} R.~S.,  {Glover} S. C.~O.,  2023, \mn@doi [\araa] {10.1146/annurev-astro-071221-053453}, \href {https://ui.adsabs.harvard.edu/abs/2023ARA&A..61...65K} {61, 65}

\bibitem[\protect\citeauthoryear{{Knollmann} \& {Knebe}}{{Knollmann} \& {Knebe}}{2009}]{Knollmann_Knebe_2009}
{Knollmann} S.~R.,  {Knebe} A.,  2009, \mn@doi [\apjs] {10.1088/0067-0049/182/2/608}, \href {https://ui.adsabs.harvard.edu/abs/2009ApJS..182..608K} {182, 608}

\bibitem[\protect\citeauthoryear{{Kozyreva} et~al.,}{{Kozyreva} et~al.}{2017}]{Kozyreva_2017}
{Kozyreva} A.,  et~al., 2017, \mn@doi [\mnras] {10.1093/mnras/stw2562}, \href {https://ui.adsabs.harvard.edu/abs/2017MNRAS.464.2854K} {464, 2854}

\bibitem[\protect\citeauthoryear{{Kudritzki} \& {Puls}}{{Kudritzki} \& {Puls}}{2000}]{Kudritzki_Puls_2000}
{Kudritzki} R.-P.,  {Puls} J.,  2000, \mn@doi [\araa] {10.1146/annurev.astro.38.1.613}, \href {https://ui.adsabs.harvard.edu/abs/2000ARA&A..38..613K} {38, 613}

\bibitem[\protect\citeauthoryear{{Larson}}{{Larson}}{1998}]{Larson_1998}
{Larson} R.~B.,  1998, \mn@doi [\mnras] {10.1046/j.1365-8711.1998.02045.x10.48550/arXiv.astro-ph/9808145}, \href {https://ui.adsabs.harvard.edu/abs/1998MNRAS.301..569L} {301, 569}

\bibitem[\protect\citeauthoryear{{Latif}, {Whalen}  \& {Khochfar}}{{Latif} et~al.}{2022}]{Latif_2022}
{Latif} M.~A.,  {Whalen} D.,   {Khochfar} S.,  2022, \mn@doi [\apj] {10.3847/1538-4357/ac3916}, \href {https://ui.adsabs.harvard.edu/abs/2022ApJ...925...28L} {925, 28}

\bibitem[\protect\citeauthoryear{{Laureijs} et~al.,}{{Laureijs} et~al.}{2011}]{Laureijs_2011}
{Laureijs} R.,  et~al., 2011, \mn@doi [arXiv e-prints] {10.48550/arXiv.1110.3193}, \href {https://ui.adsabs.harvard.edu/abs/2011arXiv1110.3193L} {p. arXiv:1110.3193}

\bibitem[\protect\citeauthoryear{{Lazar} \& {Bromm}}{{Lazar} \& {Bromm}}{2022}]{Lazar_Bromm_2022}
{Lazar} A.,  {Bromm} V.,  2022, \mn@doi [\mnras] {10.1093/mnras/stac176}, \href {https://ui.adsabs.harvard.edu/abs/2022MNRAS.511.2505L} {511, 2505}

\bibitem[\protect\citeauthoryear{{Liu} \& {Bromm}}{{Liu} \& {Bromm}}{2020}]{Liu_Bromm_2020_2}
{Liu} B.,  {Bromm} V.,  2020, \mn@doi [\mnras] {10.1093/mnras/staa2143}, \href {https://ui.adsabs.harvard.edu/abs/2020MNRAS.497.2839L} {497, 2839}

\bibitem[\protect\citeauthoryear{{Mackey}, {Bromm}  \& {Hernquist}}{{Mackey} et~al.}{2003}]{Mackey_2003}
{Mackey} J.,  {Bromm} V.,   {Hernquist} L.,  2003, \mn@doi [\apj] {10.1086/367613}, \href {https://ui.adsabs.harvard.edu/abs/2003ApJ...586....1M} {586, 1}

\bibitem[\protect\citeauthoryear{{Magg}, {Hartwig}, {Glover}, {Klessen}  \& {Whalen}}{{Magg} et~al.}{2016}]{Magg_2016}
{Magg} M.,  {Hartwig} T.,  {Glover} S. C.~O.,  {Klessen} R.~S.,   {Whalen} D.~J.,  2016, \mn@doi [\mnras] {10.1093/mnras/stw1882}, \href {https://ui.adsabs.harvard.edu/abs/2016MNRAS.462.3591M} {462, 3591}

\bibitem[\protect\citeauthoryear{{Magg}, {Schauer}, {Klessen}, {Glover}, {Tress}  \& {Jaura}}{{Magg} et~al.}{2022}]{Magg_2022}
{Magg} M.,  {Schauer} A. T.~P.,  {Klessen} R.~S.,  {Glover} S. C.~O.,  {Tress} R.~G.,   {Jaura} O.,  2022, \mn@doi [\apj] {10.3847/1538-4357/ac5aac}, \href {https://ui.adsabs.harvard.edu/abs/2022ApJ...929..119M} {929, 119}

\bibitem[\protect\citeauthoryear{{Maio}, {Ciardi}, {Dolag}, {Tornatore}  \& {Khochfar}}{{Maio} et~al.}{2010}]{Maio_2010}
{Maio} U.,  {Ciardi} B.,  {Dolag} K.,  {Tornatore} L.,   {Khochfar} S.,  2010, \mn@doi [\mnras] {10.1111/j.1365-2966.2010.17003.x}, \href {https://ui.adsabs.harvard.edu/abs/2010MNRAS.407.1003M} {407, 1003}

\bibitem[\protect\citeauthoryear{{Maio}, {Petkova}, {De Lucia}  \& {Borgani}}{{Maio} et~al.}{2016}]{Maio_2016}
{Maio} U.,  {Petkova} M.,  {De Lucia} G.,   {Borgani} S.,  2016, \mn@doi [\mnras] {10.1093/mnras/stw1196}, \href {https://ui.adsabs.harvard.edu/abs/2016MNRAS.460.3733M} {460, 3733}

\bibitem[\protect\citeauthoryear{{Maiolino} et~al.,}{{Maiolino} et~al.}{2023}]{Maiolino_2023}
{Maiolino} R.,  et~al., 2023, \mn@doi [arXiv e-prints] {10.48550/arXiv.2306.00953}, \href {https://ui.adsabs.harvard.edu/abs/2023arXiv230600953M} {p. arXiv:2306.00953}

\bibitem[\protect\citeauthoryear{{Marigo}, {Girardi}, {Chiosi}  \& {Wood}}{{Marigo} et~al.}{2001}]{Marigo_2001}
{Marigo} P.,  {Girardi} L.,  {Chiosi} C.,   {Wood} P.~R.,  2001, \mn@doi [\aap] {10.1051/0004-6361:20010309}, \href {https://ui.adsabs.harvard.edu/abs/2001A&A...371..152M} {371, 152}

\bibitem[\protect\citeauthoryear{{Mas-Ribas}, {Dijkstra}  \& {Forero-Romero}}{{Mas-Ribas} et~al.}{2016}]{Mas-Ribas_2016}
{Mas-Ribas} L.,  {Dijkstra} M.,   {Forero-Romero} J.~E.,  2016, \mn@doi [\apj] {10.3847/1538-4357/833/1/65}, \href {https://ui.adsabs.harvard.edu/abs/2016ApJ...833...65M} {833, 65}

\bibitem[\protect\citeauthoryear{{Miralda-Escud{\'e}} \& {Rees}}{{Miralda-Escud{\'e}} \& {Rees}}{1997}]{Miralda-Escude_Rees_1997}
{Miralda-Escud{\'e}} J.,  {Rees} M.~J.,  1997, \mn@doi [\apjl] {10.1086/310550}, \href {https://ui.adsabs.harvard.edu/abs/1997ApJ...478L..57M} {478, L57}

\bibitem[\protect\citeauthoryear{{Moriya}, {Tominaga}, {Tanaka}, {Maeda}  \& {Nomoto}}{{Moriya} et~al.}{2010}]{Moriya_2010}
{Moriya} T.,  {Tominaga} N.,  {Tanaka} M.,  {Maeda} K.,   {Nomoto} K.,  2010, \mn@doi [\apjl] {10.1088/2041-8205/717/2/L83}, \href {https://ui.adsabs.harvard.edu/abs/2010ApJ...717L..83M} {717, L83}

\bibitem[\protect\citeauthoryear{{Moriya} et~al.,}{{Moriya} et~al.}{2022a}]{Moriya_2022_Euclid}
{Moriya} T.~J.,  et~al., 2022a, \mn@doi [\aap] {10.1051/0004-6361/202243810}, \href {https://ui.adsabs.harvard.edu/abs/2022A&A...666A.157M} {666, A157}

\bibitem[\protect\citeauthoryear{{Moriya}, {Quimby}  \& {Robertson}}{{Moriya} et~al.}{2022b}]{Moriya_2022_RST}
{Moriya} T.~J.,  {Quimby} R.~M.,   {Robertson} B.~E.,  2022b, \mn@doi [\apj] {10.3847/1538-4357/ac415e}, \href {https://ui.adsabs.harvard.edu/abs/2022ApJ...925..211M} {925, 211}

\bibitem[\protect\citeauthoryear{{Nakajima} \& {Maiolino}}{{Nakajima} \& {Maiolino}}{2022}]{Nakajima_Maiolino_2022}
{Nakajima} K.,  {Maiolino} R.,  2022, \mn@doi [\mnras] {10.1093/mnras/stac1242}, \href {https://ui.adsabs.harvard.edu/abs/2022MNRAS.513.5134N} {513, 5134}

\bibitem[\protect\citeauthoryear{{Natarajan}, {Tan}  \& {O'Shea}}{{Natarajan} et~al.}{2009}]{Natarajan_2009}
{Natarajan} A.,  {Tan} J.~C.,   {O'Shea} B.~W.,  2009, \mn@doi [\apj] {10.1088/0004-637X/692/1/574}, \href {https://ui.adsabs.harvard.edu/abs/2009ApJ...692..574N} {692, 574}

\bibitem[\protect\citeauthoryear{{Olivier}, {Berg}, {Chisholm}, {Erb}, {Pogge}  \& {Skillman}}{{Olivier} et~al.}{2022}]{Olivier_2022}
{Olivier} G.~M.,  {Berg} D.~A.,  {Chisholm} J.,  {Erb} D.~K.,  {Pogge} R.~W.,   {Skillman} E.~D.,  2022, \mn@doi [\apj] {10.3847/1538-4357/ac8f2c}, \href {https://ui.adsabs.harvard.edu/abs/2022ApJ...938...16O} {938, 16}

\bibitem[\protect\citeauthoryear{{Omukai}, {Tsuribe}, {Schneider}  \& {Ferrara}}{{Omukai} et~al.}{2005}]{Omukai_2005}
{Omukai} K.,  {Tsuribe} T.,  {Schneider} R.,   {Ferrara} A.,  2005, \mn@doi [\apj] {10.1086/429955}, \href {https://ui.adsabs.harvard.edu/abs/2005ApJ...626..627O} {626, 627}

\bibitem[\protect\citeauthoryear{{Pallottini}, {Ferrara}, {Gallerani}, {Salvadori}  \& {D'Odorico}}{{Pallottini} et~al.}{2014}]{Pallottini_2014}
{Pallottini} A.,  {Ferrara} A.,  {Gallerani} S.,  {Salvadori} S.,   {D'Odorico} V.,  2014, \mn@doi [\mnras] {10.1093/mnras/stu451}, \href {https://ui.adsabs.harvard.edu/abs/2014MNRAS.440.2498P} {440, 2498}

\bibitem[\protect\citeauthoryear{{Pan} \& {Loeb}}{{Pan} \& {Loeb}}{2013}]{Pan_Loeb_2013}
{Pan} T.,  {Loeb} A.,  2013, \mn@doi [\mnras] {10.1093/mnrasl/slt089}, \href {https://ui.adsabs.harvard.edu/abs/2013MNRAS.435L..33P} {435, L33}

\bibitem[\protect\citeauthoryear{{Pan}, {Kasen}  \& {Loeb}}{{Pan} et~al.}{2012}]{Pan_2012_PISN}
{Pan} T.,  {Kasen} D.,   {Loeb} A.,  2012, \mn@doi [\mnras] {10.1111/j.1365-2966.2012.20837.x}, \href {https://ui.adsabs.harvard.edu/abs/2012MNRAS.422.2701P} {422, 2701}

\bibitem[\protect\citeauthoryear{{Pirzkal}, {Finkelstein}, {Papovich}  \& {Ngdeep Team}}{{Pirzkal} et~al.}{2023}]{Pirzkal_2023}
{Pirzkal} N.,  {Finkelstein} S.,  {Papovich} C.,   {Ngdeep Team} 2023, in American Astronomical Society Meeting Abstracts. p. 455.04

\bibitem[\protect\citeauthoryear{{Planck Collaboration} et~al.,}{{Planck Collaboration} et~al.}{2016}]{Planck_2015}
{Planck Collaboration} et~al., 2016, \mn@doi [\aap] {10.1051/0004-6361/201525830}, \href {https://ui.adsabs.harvard.edu/abs/2016A&A...594A..13P} {594, A13}

\bibitem[\protect\citeauthoryear{{Raiter}, {Schaerer}  \& {Fosbury}}{{Raiter} et~al.}{2010}]{Raiter_2010}
{Raiter} A.,  {Schaerer} D.,   {Fosbury} R.~A.~E.,  2010, \mn@doi [\aap] {10.1051/0004-6361/201015236}, \href {https://ui.adsabs.harvard.edu/abs/2010A&A...523A..64R} {523, A64}

\bibitem[\protect\citeauthoryear{{Rakavy} \& {Shaviv}}{{Rakavy} \& {Shaviv}}{1967}]{Rakavy_Shaviv_1967}
{Rakavy} G.,  {Shaviv} G.,  1967, \mn@doi [\apj] {10.1086/149204}, \href {https://ui.adsabs.harvard.edu/abs/1967ApJ...148..803R} {148, 803}

\bibitem[\protect\citeauthoryear{{Rees} \& {Ostriker}}{{Rees} \& {Ostriker}}{1977}]{Rees_Ostriker_1977}
{Rees} M.~J.,  {Ostriker} J.~P.,  1977, \mnras, \href {https://ui.adsabs.harvard.edu/abs/1977MNRAS.179..541R} {179, 541}

\bibitem[\protect\citeauthoryear{{Reg{\H{o}}s}, {Vink{\'o}}  \& {Ziegler}}{{Reg{\H{o}}s} et~al.}{2020}]{Regos_2020}
{Reg{\H{o}}s} E.,  {Vink{\'o}} J.,   {Ziegler} B.~L.,  2020, \mn@doi [\apj] {10.3847/1538-4357/ab8636}, \href {https://ui.adsabs.harvard.edu/abs/2020ApJ...894...94R} {894, 94}

\bibitem[\protect\citeauthoryear{{Riaz}, {Hartwig}  \& {Latif}}{{Riaz} et~al.}{2022}]{Riaz_2022}
{Riaz} S.,  {Hartwig} T.,   {Latif} M.~A.,  2022, \mn@doi [\apjl] {10.3847/2041-8213/ac8ea6}, \href {https://ui.adsabs.harvard.edu/abs/2022ApJ...937L...6R} {937, L6}

\bibitem[\protect\citeauthoryear{{Rossi}, {Salvadori}  \& {Sk{\'u}lad{\'o}ttir}}{{Rossi} et~al.}{2021}]{Rossi_2021}
{Rossi} M.,  {Salvadori} S.,   {Sk{\'u}lad{\'o}ttir} {\'A}.,  2021, \mn@doi [\mnras] {10.1093/mnras/stab82110.48550/arXiv.2103.09834}, \href {https://ui.adsabs.harvard.edu/abs/2021MNRAS.503.6026R} {503, 6026}

\bibitem[\protect\citeauthoryear{{Salpeter}}{{Salpeter}}{1955}]{Salpeter_1955}
{Salpeter} E.~E.,  1955, \mn@doi [\apj] {10.1086/145971}, \href {https://ui.adsabs.harvard.edu/abs/1955ApJ...121..161S} {121, 161}

\bibitem[\protect\citeauthoryear{{Salvadori}, {Schneider}  \& {Ferrara}}{{Salvadori} et~al.}{2007}]{Salvadori_2007}
{Salvadori} S.,  {Schneider} R.,   {Ferrara} A.,  2007, \mn@doi [\mnras] {10.1111/j.1365-2966.2007.12133.x}, \href {https://ui.adsabs.harvard.edu/abs/2007MNRAS.381..647S} {381, 647}

\bibitem[\protect\citeauthoryear{{Salvadori}, {Ferrara}  \& {Schneider}}{{Salvadori} et~al.}{2008}]{Salvadori_2008}
{Salvadori} S.,  {Ferrara} A.,   {Schneider} R.,  2008, \mn@doi [\mnras] {10.1111/j.1365-2966.2008.13035.x}, \href {https://ui.adsabs.harvard.edu/abs/2008MNRAS.386..348S} {386, 348}

\bibitem[\protect\citeauthoryear{{Salvadori}, {Bonifacio}, {Caffau}, {Korotin}, {Andreevsky}, {Spite}  \& {Sk{\'u}lad{\'o}ttir}}{{Salvadori} et~al.}{2019}]{Salvadori_2019}
{Salvadori} S.,  {Bonifacio} P.,  {Caffau} E.,  {Korotin} S.,  {Andreevsky} S.,  {Spite} M.,   {Sk{\'u}lad{\'o}ttir} {\'A}.,  2019, \mn@doi [\mnras] {10.1093/mnras/stz1464}, \href {https://ui.adsabs.harvard.edu/abs/2019MNRAS.487.4261S} {487, 4261}

\bibitem[\protect\citeauthoryear{{Sarmento} \& {Scannapieco}}{{Sarmento} \& {Scannapieco}}{2022}]{Sarmento_Scannapieco_2022}
{Sarmento} R.,  {Scannapieco} E.,  2022, \mn@doi [\apj] {10.3847/1538-4357/ac815c}, \href {https://ui.adsabs.harvard.edu/abs/2022ApJ...935..174S} {935, 174}

\bibitem[\protect\citeauthoryear{{Sarmento}, {Scannapieco}  \& {Pan}}{{Sarmento} et~al.}{2016}]{Sarmento_2016}
{Sarmento} R.~J.,  {Scannapieco} E.,   {Pan} L.,  2016, in American Astronomical Society Meeting Abstracts \#228. p. 319.11

\bibitem[\protect\citeauthoryear{{Sarmento}, {Scannapieco}  \& {Pan}}{{Sarmento} et~al.}{2017}]{Sarmento_2017}
{Sarmento} R.,  {Scannapieco} E.,   {Pan} L.,  2017, \mn@doi [\apj] {10.3847/1538-4357/834/1/23}, \href {https://ui.adsabs.harvard.edu/abs/2017ApJ...834...23S} {834, 23}

\bibitem[\protect\citeauthoryear{{Sarmento}, {Scannapieco}  \& {Cohen}}{{Sarmento} et~al.}{2018}]{Sarmento_2018}
{Sarmento} R.,  {Scannapieco} E.,   {Cohen} S.,  2018, \mn@doi [\apj] {10.3847/1538-4357/aa989a}, \href {https://ui.adsabs.harvard.edu/abs/2018ApJ...854...75S} {854, 75}

\bibitem[\protect\citeauthoryear{{Scannapieco}, {Madau}, {Woosley}, {Heger}  \& {Ferrara}}{{Scannapieco} et~al.}{2005}]{Scannapieco_2005}
{Scannapieco} E.,  {Madau} P.,  {Woosley} S.,  {Heger} A.,   {Ferrara} A.,  2005, \mn@doi [\apj] {10.1086/444450}, \href {https://ui.adsabs.harvard.edu/abs/2005ApJ...633.1031S} {633, 1031}

\bibitem[\protect\citeauthoryear{{Schaerer}}{{Schaerer}}{2002}]{Schaerer_2002}
{Schaerer} D.,  2002, \mn@doi [\aap] {10.1051/0004-6361:20011619}, \href {https://ui.adsabs.harvard.edu/abs/2002A&A...382...28S} {382, 28}

\bibitem[\protect\citeauthoryear{{Schaerer}}{{Schaerer}}{2003}]{Schaerer_2003}
{Schaerer} D.,  2003, \mn@doi [\aap] {10.1051/0004-6361:20021525}, \href {https://ui.adsabs.harvard.edu/abs/2003A&A...397..527S} {397, 527}

\bibitem[\protect\citeauthoryear{{Schauer}, {Bromm}, {Drory}  \& {Boylan-Kolchin}}{{Schauer} et~al.}{2022}]{Schauer_2022}
{Schauer} A. T.~P.,  {Bromm} V.,  {Drory} N.,   {Boylan-Kolchin} M.,  2022, \mn@doi [\apjl] {10.3847/2041-8213/ac7f9a}, \href {https://ui.adsabs.harvard.edu/abs/2022ApJ...934L...6S} {934, L6}

\bibitem[\protect\citeauthoryear{{Schneider}, {Omukai}, {Bianchi}  \& {Valiante}}{{Schneider} et~al.}{2012a}]{Schneider_2012_dustToGas}
{Schneider} R.,  {Omukai} K.,  {Bianchi} S.,   {Valiante} R.,  2012a, \mn@doi [\mnras] {10.1111/j.1365-2966.2011.19818.x}, \href {https://ui.adsabs.harvard.edu/abs/2012MNRAS.419.1566S} {419, 1566}

\bibitem[\protect\citeauthoryear{{Schneider}, {Omukai}, {Limongi}, {Ferrara}, {Salvaterra}, {Chieffi}  \& {Bianchi}}{{Schneider} et~al.}{2012b}]{Schneider_2012_dustSDSSstar}
{Schneider} R.,  {Omukai} K.,  {Limongi} M.,  {Ferrara} A.,  {Salvaterra} R.,  {Chieffi} A.,   {Bianchi} S.,  2012b, \mn@doi [\mnras] {10.1111/j.1745-3933.2012.01257.x}, \href {https://ui.adsabs.harvard.edu/abs/2012MNRAS.423L..60S} {423, L60}

\bibitem[\protect\citeauthoryear{{Skinner} \& {Wise}}{{Skinner} \& {Wise}}{2020}]{Skinner_Wise_2020}
{Skinner} D.,  {Wise} J.~H.,  2020, \mn@doi [\mnras] {10.1093/mnras/staa139}, \href {https://ui.adsabs.harvard.edu/abs/2020MNRAS.492.4386S} {492, 4386}

\bibitem[\protect\citeauthoryear{{Spolyar}, {Freese}  \& {Gondolo}}{{Spolyar} et~al.}{2008}]{Spoylar_2008}
{Spolyar} D.,  {Freese} K.,   {Gondolo} P.,  2008, \mn@doi [\prl] {10.1103/PhysRevLett.100.051101}, \href {https://ui.adsabs.harvard.edu/abs/2008PhRvL.100e1101S} {100, 051101}

\bibitem[\protect\citeauthoryear{{Springel} \& {Hernquist}}{{Springel} \& {Hernquist}}{2003}]{Springel_Hernquist_2003}
{Springel} V.,  {Hernquist} L.,  2003, \mn@doi [\mnras] {10.1046/j.1365-8711.2003.06206.x}, \href {https://ui.adsabs.harvard.edu/abs/2003MNRAS.339..289S} {339, 289}

\bibitem[\protect\citeauthoryear{{Stacy}, {Bromm}  \& {Lee}}{{Stacy} et~al.}{2016}]{Stacy_2016}
{Stacy} A.,  {Bromm} V.,   {Lee} A.~T.,  2016, \mn@doi [\mnras] {10.1093/mnras/stw1728}, \href {https://ui.adsabs.harvard.edu/abs/2016MNRAS.462.1307S} {462, 1307}

\bibitem[\protect\citeauthoryear{{Su}, {Hopkins}, {Hayward}, {Faucher-Gigu{\`e}re}, {Kere{\v{s}}}, {Ma}  \& {Robles}}{{Su} et~al.}{2017}]{Su_2017}
{Su} K.-Y.,  {Hopkins} P.~F.,  {Hayward} C.~C.,  {Faucher-Gigu{\`e}re} C.-A.,  {Kere{\v{s}}} D.,  {Ma} X.,   {Robles} V.~H.,  2017, \mn@doi [\mnras] {10.1093/mnras/stx1463}, \href {https://ui.adsabs.harvard.edu/abs/2017MNRAS.471..144S} {471, 144}

\bibitem[\protect\citeauthoryear{{Sugimura}, {Matsumoto}, {Hosokawa}, {Hirano}  \& {Omukai}}{{Sugimura} et~al.}{2020}]{Sugimura_2020}
{Sugimura} K.,  {Matsumoto} T.,  {Hosokawa} T.,  {Hirano} S.,   {Omukai} K.,  2020, \mn@doi [\apjl] {10.3847/2041-8213/ab7d37}, \href {https://ui.adsabs.harvard.edu/abs/2020ApJ...892L..14S} {892, L14}

\bibitem[\protect\citeauthoryear{{Susa}, {Hasegawa}  \& {Tominaga}}{{Susa} et~al.}{2014}]{Susa_2014}
{Susa} H.,  {Hasegawa} K.,   {Tominaga} N.,  2014, \mn@doi [\apj] {10.1088/0004-637X/792/1/32}, \href {https://ui.adsabs.harvard.edu/abs/2014ApJ...792...32S} {792, 32}

\bibitem[\protect\citeauthoryear{{Tanaka}, {Moriya}  \& {Yoshida}}{{Tanaka} et~al.}{2013}]{Tanaka_2013}
{Tanaka} M.,  {Moriya} T.~J.,   {Yoshida} N.,  2013, \mn@doi [\mnras] {10.1093/mnras/stt1469}, \href {https://ui.adsabs.harvard.edu/abs/2013MNRAS.435.2483T} {435, 2483}

\bibitem[\protect\citeauthoryear{{Thielemann} et~al.,}{{Thielemann} et~al.}{2003}]{Thielemann_2003}
{Thielemann} F.~K.,  et~al., 2003, \mn@doi [\nphysa] {10.1016/S0375-9474(03)00704-8}, \href {https://ui.adsabs.harvard.edu/abs/2003NuPhA.718..139T} {718, 139}

\bibitem[\protect\citeauthoryear{{Tornatore}, {Ferrara}  \& {Schneider}}{{Tornatore} et~al.}{2007a}]{Tornatore_2007_PopIII}
{Tornatore} L.,  {Ferrara} A.,   {Schneider} R.,  2007a, \mn@doi [\mnras] {10.1111/j.1365-2966.2007.12215.x}, \href {https://ui.adsabs.harvard.edu/abs/2007MNRAS.382..945T} {382, 945}

\bibitem[\protect\citeauthoryear{{Tornatore}, {Borgani}, {Dolag}  \& {Matteucci}}{{Tornatore} et~al.}{2007b}]{Tornatore_2007_chemicalFeedback}
{Tornatore} L.,  {Borgani} S.,  {Dolag} K.,   {Matteucci} F.,  2007b, \mn@doi [\mnras] {10.1111/j.1365-2966.2007.12070.x}, \href {https://ui.adsabs.harvard.edu/abs/2007MNRAS.382.1050T} {382, 1050}

\bibitem[\protect\citeauthoryear{{Torres}, {Stefanik}  \& {Latham}}{{Torres} et~al.}{1997}]{Torres_1997}
{Torres} G.,  {Stefanik} R.~P.,   {Latham} D.~W.,  1997, \mn@doi [\apj] {10.1086/304422}, \href {https://ui.adsabs.harvard.edu/abs/1997ApJ...485..167T} {485, 167}

\bibitem[\protect\citeauthoryear{{Treu} et~al.,}{{Treu} et~al.}{2017}]{Treu_2017}
{Treu} T.~L.,  et~al., 2017, {Through the Looking GLASS: A JWST Exploration of Galaxy Formation and Evolution from Cosmic Dawn to Present Day}, JWST Proposal ID 1324. Cycle 0 Early Release Science

\bibitem[\protect\citeauthoryear{{Treu} et~al.,}{{Treu} et~al.}{2022}]{Treu_2022}
{Treu} T.,  et~al., 2022, \mn@doi [\apj] {10.3847/1538-4357/ac8158}, \href {https://ui.adsabs.harvard.edu/abs/2022ApJ...935..110T} {935, 110}

\bibitem[\protect\citeauthoryear{{Trinca}, {Schneider}, {Valiante}, {Graziani}, {Ferrotti}, {Omukai}  \& {Chon}}{{Trinca} et~al.}{2023}]{Trinca_2023}
{Trinca} A.,  {Schneider} R.,  {Valiante} R.,  {Graziani} L.,  {Ferrotti} A.,  {Omukai} K.,   {Chon} S.,  2023, \mn@doi [arXiv e-prints] {10.48550/arXiv.2305.04944}, \href {https://ui.adsabs.harvard.edu/abs/2023arXiv230504944T} {p. arXiv:2305.04944}

\bibitem[\protect\citeauthoryear{{Trussler} et~al.,}{{Trussler} et~al.}{2023}]{Trussler_2023}
{Trussler} J. A.~A.,  et~al., 2023, \mn@doi [\mnras] {10.1093/mnras/stad2553}, \href {https://ui.adsabs.harvard.edu/abs/2023MNRAS.525.5328T} {525, 5328}

\bibitem[\protect\citeauthoryear{{Tumlinson} \& {Shull}}{{Tumlinson} \& {Shull}}{2000}]{Tumlinson_Shull_2000}
{Tumlinson} J.,  {Shull} J.~M.,  2000, \mn@doi [\apjl] {10.1086/312432}, \href {https://ui.adsabs.harvard.edu/abs/2000ApJ...528L..65T} {528, L65}

\bibitem[\protect\citeauthoryear{{Tumlinson}, {Giroux}  \& {Shull}}{{Tumlinson} et~al.}{2001}]{Tumlinson_2001}
{Tumlinson} J.,  {Giroux} M.~L.,   {Shull} J.~M.,  2001, \mn@doi [\apjl] {10.1086/319477}, \href {https://ui.adsabs.harvard.edu/abs/2001ApJ...550L...1T} {550, L1}

\bibitem[\protect\citeauthoryear{{Vanni}, {Salvadori}  \& {Sk{\'u}lad{\'o}ttir}}{{Vanni} et~al.}{2023}]{Vanni_2023}
{Vanni} I.,  {Salvadori} S.,   {Sk{\'u}lad{\'o}ttir} {\'A}.,  2023, \mn@doi [arXiv e-prints] {10.48550/arXiv.2305.02358}, \href {https://ui.adsabs.harvard.edu/abs/2023arXiv230502358V} {p. arXiv:2305.02358}

\bibitem[\protect\citeauthoryear{{Vanzella} et~al.,}{{Vanzella} et~al.}{2020}]{Vanzella_2020}
{Vanzella} E.,  et~al., 2020, \mn@doi [\mnras] {10.1093/mnrasl/slaa041}, \href {https://ui.adsabs.harvard.edu/abs/2020MNRAS.494L..81V} {494, L81}

\bibitem[\protect\citeauthoryear{{Vanzella} et~al.,}{{Vanzella} et~al.}{2023}]{Vanzella_2023}
{Vanzella} E.,  et~al., 2023, \mn@doi [\aap] {10.1051/0004-6361/202346981}, \href {https://ui.adsabs.harvard.edu/abs/2023A&A...678A.173V} {678, A173}

\bibitem[\protect\citeauthoryear{{Venditti}, {Graziani}, {Schneider}, {Pentericci}, {Di Cesare}, {Maio}  \& {Omukai}}{{Venditti} et~al.}{2023}]{Venditti_2023}
{Venditti} A.,  {Graziani} L.,  {Schneider} R.,  {Pentericci} L.,  {Di Cesare} C.,  {Maio} U.,   {Omukai} K.,  2023, \mn@doi [\mnras] {10.1093/mnras/stad1201}, \href {https://ui.adsabs.harvard.edu/abs/2023MNRAS.522.3809V} {522, 3809}

\bibitem[\protect\citeauthoryear{{Visbal}, {Bryan}  \& {Haiman}}{{Visbal} et~al.}{2020}]{Visbal_2020}
{Visbal} E.,  {Bryan} G.~L.,   {Haiman} Z.,  2020, \mn@doi [\apj] {10.3847/1538-4357/ab994e}, \href {https://ui.adsabs.harvard.edu/abs/2020ApJ...897...95V} {897, 95}

\bibitem[\protect\citeauthoryear{{Wang}, {Bromm}, {Greif}, {Stacy}, {Dai}, {Loeb}  \& {Cheng}}{{Wang} et~al.}{2012}]{Wang_2012}
{Wang} F.~Y.,  {Bromm} V.,  {Greif} T.~H.,  {Stacy} A.,  {Dai} Z.~G.,  {Loeb} A.,   {Cheng} K.~S.,  2012, \mn@doi [\apj] {10.1088/0004-637X/760/1/27}, \href {https://ui.adsabs.harvard.edu/abs/2012ApJ...760...27W} {760, 27}

\bibitem[\protect\citeauthoryear{{Wang} et~al.,}{{Wang} et~al.}{2017}]{Wang_2017}
{Wang} L.,  et~al., 2017, \mn@doi [arXiv e-prints] {10.48550/arXiv.1710.07005}, \href {https://ui.adsabs.harvard.edu/abs/2017arXiv171007005W} {p. arXiv:1710.07005}

\bibitem[\protect\citeauthoryear{{Wang} et~al.,}{{Wang} et~al.}{2022}]{Wang_2022}
{Wang} X.,  et~al., 2022, arXiv e-prints, \href {https://ui.adsabs.harvard.edu/abs/2022arXiv221204476W} {p. arXiv:2212.04476}

\bibitem[\protect\citeauthoryear{{Weaver} et~al.,}{{Weaver} et~al.}{2023}]{Weaver_2023}
{Weaver} J.~R.,  et~al., 2023, \mn@doi [arXiv e-prints] {10.48550/arXiv.2301.02671}, \href {https://ui.adsabs.harvard.edu/abs/2023arXiv230102671W} {p. arXiv:2301.02671}

\bibitem[\protect\citeauthoryear{{Whalen} et~al.,}{{Whalen} et~al.}{2013}]{Whalen_2013}
{Whalen} D.~J.,  et~al., 2013, \mn@doi [\apj] {10.1088/0004-637X/777/2/110}, \href {https://ui.adsabs.harvard.edu/abs/2013ApJ...777..110W} {777, 110}

\bibitem[\protect\citeauthoryear{{Whalen} et~al.,}{{Whalen} et~al.}{2014}]{Whalen_2014}
{Whalen} D.~J.,  et~al., 2014, \mn@doi [\apj] {10.1088/0004-637X/797/1/9}, \href {https://ui.adsabs.harvard.edu/abs/2014ApJ...797....9W} {797, 9}

\bibitem[\protect\citeauthoryear{{Williams} et~al.,}{{Williams} et~al.}{2021}]{Williams_2021}
{Williams} C.~C.,  et~al., 2021, {PANORAMIC - A Pure Parallel Wide Area Legacy Imaging Survey at 1-5 Micron}, JWST Proposal. Cycle 1, ID. \#2514

\bibitem[\protect\citeauthoryear{{Wise} \& {Abel}}{{Wise} \& {Abel}}{2005}]{Wise_Abel_2005}
{Wise} J.~H.,  {Abel} T.,  2005, \mn@doi [\apj] {10.1086/430434}, \href {https://ui.adsabs.harvard.edu/abs/2005ApJ...629..615W} {629, 615}

\bibitem[\protect\citeauthoryear{{Wise}, {Turk}, {Norman}  \& {Abel}}{{Wise} et~al.}{2012}]{Wise_2012_PMEnrichment}
{Wise} J.~H.,  {Turk} M.~J.,  {Norman} M.~L.,   {Abel} T.,  2012, \mn@doi [\apj] {10.1088/0004-637X/745/1/50}, \href {https://ui.adsabs.harvard.edu/abs/2012ApJ...745...50W} {745, 50}

\bibitem[\protect\citeauthoryear{{Woosley} \& {Weaver}}{{Woosley} \& {Weaver}}{1995}]{Woosley_Weaver_1995}
{Woosley} S.~E.,  {Weaver} T.~A.,  1995, \mn@doi [\apjs] {10.1086/192237}, \href {https://ui.adsabs.harvard.edu/abs/1995ApJS..101..181W} {101, 181}

\bibitem[\protect\citeauthoryear{{Woosley}, {Blinnikov}  \& {Heger}}{{Woosley} et~al.}{2007}]{Woosley_2007}
{Woosley} S.~E.,  {Blinnikov} S.,   {Heger} A.,  2007, \mn@doi [\nat] {10.1038/nature06333}, \href {https://ui.adsabs.harvard.edu/abs/2007Natur.450..390W} {450, 390}

\bibitem[\protect\citeauthoryear{{Xing} et~al.,}{{Xing} et~al.}{2023}]{Xing_2023}
{Xing} Q.-F.,  et~al., 2023, \mn@doi [\nat] {10.1038/s41586-023-06028-1}, \href {https://ui.adsabs.harvard.edu/abs/2023Natur.618..712X} {618, 712}

\bibitem[\protect\citeauthoryear{{Xu}, {Norman}, {O'Shea}  \& {Wise}}{{Xu} et~al.}{2016}]{Xu_2016_latePopIII}
{Xu} H.,  {Norman} M.~L.,  {O'Shea} B.~W.,   {Wise} J.~H.,  2016, \mn@doi [\apj] {10.3847/0004-637X/823/2/140}, \href {https://ui.adsabs.harvard.edu/abs/2016ApJ...823..140X} {823, 140}

\bibitem[\protect\citeauthoryear{{Yoon}, {Iocco}  \& {Akiyama}}{{Yoon} et~al.}{2008}]{Yoon_2008}
{Yoon} S.-C.,  {Iocco} F.,   {Akiyama} S.,  2008, \mn@doi [\apjl] {10.1086/593976}, \href {https://ui.adsabs.harvard.edu/abs/2008ApJ...688L...1Y} {688, L1}

\bibitem[\protect\citeauthoryear{{Yoshii}, {Sameshima}, {Tsujimoto}, {Shigeyama}, {Beers}  \& {Peterson}}{{Yoshii} et~al.}{2022}]{Yoshii_2022}
{Yoshii} Y.,  {Sameshima} H.,  {Tsujimoto} T.,  {Shigeyama} T.,  {Beers} T.~C.,   {Peterson} B.~A.,  2022, \mn@doi [\apj] {10.3847/1538-4357/ac8163}, \href {https://ui.adsabs.harvard.edu/abs/2022ApJ...937...61Y} {937, 61}

\bibitem[\protect\citeauthoryear{{Yung}, {Somerville}, {Finkelstein}, {Wilkins}  \& {Gardner}}{{Yung} et~al.}{2023}]{Yung_2023}
{Yung} L.~Y.~A.,  {Somerville} R.~S.,  {Finkelstein} S.~L.,  {Wilkins} S.~M.,   {Gardner} J.~P.,  2023, \mn@doi [arXiv e-prints] {10.48550/arXiv.2304.04348}, \href {https://ui.adsabs.harvard.edu/abs/2023arXiv230404348Y} {p. arXiv:2304.04348}

\bibitem[\protect\citeauthoryear{Zackrisson, Rydberg, Schaerer, Östlin  \& Tuli}{Zackrisson et~al.}{2011}]{Zackrisson_2011}
Zackrisson E.,  Rydberg C.-E.,  Schaerer D.,  Östlin G.,   Tuli M.,  2011, \apj, \href {https://ui.adsabs.harvard.edu/abs/2011ApJ...740...13Z/abstract} {740, 13}

\bibitem[\protect\citeauthoryear{{de Bennassuti}, {Schneider}, {Valiante}  \& {Salvadori}}{{de Bennassuti} et~al.}{2014}]{deBennassuti_2014}
{de Bennassuti} M.,  {Schneider} R.,  {Valiante} R.,   {Salvadori} S.,  2014, \mn@doi [\mnras] {10.1093/mnras/stu1962}, \href {https://ui.adsabs.harvard.edu/abs/2014MNRAS.445.3039D} {445, 3039}

\bibitem[\protect\citeauthoryear{{de Bennassuti}, {Salvadori}, {Schneider}, {Valiante}  \& {Omukai}}{{de Bennassuti} et~al.}{2017}]{deBennassuti_2017}
{de Bennassuti} M.,  {Salvadori} S.,  {Schneider} R.,  {Valiante} R.,   {Omukai} K.,  2017, \mn@doi [\mnras] {10.1093/mnras/stw2687}, \href {https://ui.adsabs.harvard.edu/abs/2017MNRAS.465..926D} {465, 926}

\bibitem[\protect\citeauthoryear{{de Souza}, {Ishida}, {Whalen}, {Johnson}  \& {Ferrara}}{{de Souza} et~al.}{2014}]{deSouza_2014}
{de Souza} R.~S.,  {Ishida} E.~E.~O.,  {Whalen} D.~J.,  {Johnson} J.~L.,   {Ferrara} A.,  2014, \mn@doi [\mnras] {10.1093/mnras/stu984}, \href {https://ui.adsabs.harvard.edu/abs/2014MNRAS.442.1640D} {442, 1640}

\bibitem[\protect\citeauthoryear{{van den Hoek} \& {Groenewegen}}{{van den Hoek} \& {Groenewegen}}{1997}]{vanDenHoek_Groenewegen_1997}
{van den Hoek} L.~B.,  {Groenewegen} M.~A.~T.,  1997, \mn@doi [\aaps] {10.1051/aas:1997162}, \href {https://ui.adsabs.harvard.edu/abs/1997A&AS..123..305V} {123, 305}

\makeatother
\end{thebibliography}

\label{lastpage}
\end{document}